\documentclass[fleqn,usenatbib]{mnras}

\usepackage{newtxtext,newtxmath}
\usepackage[T1]{fontenc}

\DeclareRobustCommand{\VAN}[3]{#2}
\let\VANthebibliography\thebibliography
\def\thebibliography{\DeclareRobustCommand{\VAN}[3]{##3}\VANthebibliography}

\usepackage{graphicx}	
\usepackage{amsmath}	
\usepackage{booktabs}
\usepackage{subcaption}
\usepackage{siunitx}
\usepackage{floatpag}
\usepackage{float}
\usepackage[export]{adjustbox}
\usepackage{xcolor}
\usepackage{makecell}

\title [Simulations of Highly-Neutronized Ejecta in the SNR 3C 397]{Hydrodynamical Simulations Favor a Pure Deflagration Origin of the Near-Chandrasekhar Mass Supernova Remnant 3C 397}

\author[V. Mehta et al.]{
 Vrutant Mehta$^{1}$ \thanks{E-mail: vmehta2@umassd.edu},
 Jack Sullivan$^{1,2}$,
 Robert Fisher$^{1}$ \thanks{E-mail: robert.fisher@umassd.edu},
 Yuken Ohshiro$^{3}$,
 Hiroya Yamaguchi $^{3}$,
 \newauthor{
 Khanak Bhargava$^{4,5}$,
 and Sudarshan Neopane$^{6}$}
 \\
$^{1}$Department of Physics, University of Massachusetts Dartmouth, 285, Old Westport Road, North Dartmouth, 02740, USA \\
$^{2}$ Department of Physics, University of Connecticut, 196A Auditorium Rd Unit 3046, Storrs, CT 06269, USA \\
$^{3}$Department of Physics, Graduate School of Science, The University of Tokyo, 7-3-1 Hongo, Bunkyo-ku, Tokyo 113-0033, Japan \\
$^{4}$Department of Physics and Astronomy, Stony Brook University, Stony Brook, NY 11794-3800, USA \\
$^{5}$Institute for Advanced Computational Science, Stony Brook University, Stony Brook, NY 11794, USA \\
$^{6}$Department of Physics and Astronomy, University of Tennessee Knoxville, 401 Nielsen Physics Building, 1408 Circle Drive, Knoxville TN 37996-1200, USA \\
}

\date{Accepted XXX. Received YYY; in original form ZZZ}

\pubyear{2023}

\begin{document}
\label{firstpage}
\pagerange{\pageref{firstpage}--\pageref{lastpage}}
\maketitle

\begin{abstract}
Suzaku X-ray observations of the Type Ia supernova remnant (SNR) 3C 397 discovered exceptionally high mass ratios of Mn/Fe, Ni/Fe, and Cr/Fe, consistent with a near $M_{\rm Ch}$ progenitor white dwarf (WD). The Suzaku observations have established 3C 397 as our best candidate for a near-$M_{\rm Ch}$ SNR Ia, and opened the way to address additional outstanding questions about the origin and explosion mechanism of these transients. In particular, subsequent XMM-Newton observations revealed an unusually clumpy distribution of iron group elemental (IGE) abundances within the ejecta of 3C 397. In this paper, we undertake a suite of two dimensional hydrodynamical models, varying both the explosion mechanism  -- either deflagration-to-detonation (DDT), or pure deflagration --  WD progenitors, and WD progenitor metallicity, and analyze their detailed nucleosynthetic abundances and associated clumping. We find that pure deflagrations naturally give rise to clumpy distributions of neutronized species concentrated towards the outer limb of the remnant, and confirm DDTs have smoothly structured ejecta with a central concentration of neutronization. Our findings indicate that 3C 397 was most likely a pure deflagration of a high central density WD. We discuss a range of implications of these findings for the broader SN Ia progenitor problem.

\end{abstract}

\begin{keywords}
supernova remnants -- supernovae:general -- white dwarfs
\end{keywords}



\section{Introduction}



Type Ia supernovae (SNe Ia) are among the most luminous and common optical transients in the universe.  SNe Ia have important ramifications across a wide range of astrophysical domains, from galaxy formation \citep {gandhietal22} and galactic nucleosynthesis  \citep{kobayashietal20}, to the Hubble tension  \citep {reissetal16}, and the nature of dark energy \citep {maozetal14}. Yet, we still do not know the stellar progenitor systems which give rise to SNe Ia. We also do not understand  precisely how a detonation arises within these systems, and under what conditions detonations fail to initiate. These failed detonation systems will  lead to faint SNe Iax transient events with surviving hypervelocity white dwarfs (WDs) \citep {jordanetal12a, kromeretal13, foleyetal13, raddietal18, elbadryetal23}.

Galactic supernova remnants (SNRs) offer valuable insights into Type Ia supernovae (SNe Ia) through spatially-resolved X-ray observations. These observations complement optical and infrared studies of SNe Ia transients \citep {vink12}. Among SNRs, 3C 397 stands out as a uniquely rich testing ground for investigating the SN Ia progenitor problem. JAXA Suzaku X-ray observations revealed high abundance ratios of Ni/Fe, Mn/Fe, and Cr/Fe \citep {yamaguchietal15}. Such high abundance ratios of iron-group elements (IGEs) point towards a near-Chandrasekhar (near-$M_{\rm Ch}$) mass white dwarf (WD) progenitor for 3C 397, since the dense interiors of near-$M_{\rm Ch}$ WD SNe Ia undergo efficient electron captures, producing neutronized isotopes in abundance, including $^{62}$Ni,  $^{55}$Fe (which subsequently decays to $^{55}$Mn), and $^{54}$Cr. Further detailed nucleosynthesis from hydrodynamical simulations  predicted that in order to match the exceptionally high IGE abundance ratios observed in 3C 397, the WD progenitor of 3C 397 either had to have either a  high central density ($\rho_c \sim 5 - 6 \times 10^9$ g cm$^{-3}$), or represent a pure deflagration of a canonical near-$M_{\rm Ch}$  WD  ($\rho_c \sim 2 \times 10^9$ g cm$^{-3}$) often adopted in the literature \citep {daveetal17, leungnomoto18}. These predictions for a high-central density near-$M_{\rm Ch}$  WD progenitor for SNR 3C 397 gathered support in ESA XMM-Newton follow-up observations, which uncovered the existence of clumps with high Cr/Fe abundances, indicative of highly-neutronized burning \citep {oshiroetal21}. However, the clumpiness of the ejecta is in tension with the layered structure generally predicted by deflagration-to-detonation transition (DDT) models \citep {gamezoetal04}. Both the explosion mechanism as well as the structure of the WD progenitor of 3C 397 at ignition (which depends upon the initial WD mass, accretion rate, and angular momentum in the single-degenerate near-$M_{\rm Ch}$ scenario) have important implications for the SN Ia problem in general.

This  successful dialog between X-ray observers and theorists lays the foundation for the next stage of progress in understanding the origin of SNR 3C 397, and its connection to the wider SN Ia progenitor problem. While many diverse channels have been predicted to contribute to the SN Ia rate, all leading models for normal brightness SNe Ia share in common one physical process: detonation. 
In this study, we investigate two distinct explosion mechanisms in our simulations.

The first scenario we will investigate is the deflagration-to-detonation transition (DDT). The DDT has been a well-explored explosion mechanism over the past few decades in the general context of the near-$M_{\rm Ch}$ channel of SNe Ia. In the DDT scenario, a flame bubble ignites runaway deflagration burning inside the convective region of the near-$M_{\rm Ch}$ white dwarf, and  rises buoyantly until it transitions to a detonation. The nuclear energy released during the detonation is sufficient to completely disrupt the white dwarf, and typically produces a normal to bright type Ia event. 
The solar abundance of the monoisotopic element $^{55}$Mn strongly suggests that near-$M_{\rm Ch}$ SNe Ia explosions contribute at least partially to the total SNe Ia rate \citep {seitenzahletal13}. Additionally, recent JWST observations of SN 2021aefx favor a DDT in a near-$M_{\rm Ch}$ WD progenitor model \citep {derkacyetal23}.

In contrast, in the second scenario we investigate, a pure deflagration,  the  flame bubble ignites close to the center of the WD and burns through the dense core. After burning through the center of the WD, the bubble rises to the surface without a transition into detonation. Only a fraction of the total mass of the WD is burned; this model predicts that failed detonation near-$M_{\rm Ch}$ events naturally produce significantly less $^{56}$Ni and ejected mass than a DDT \citep {ropkeetal07, longetal14}. The outcome is therefore a  subluminous and more rapidly declining transient than a normal event \citep {jordanetal12b, kromeretal13, finketal14}. These predictions are broadly consistent with the class of subluminuous Iax events \citep {foleyetal13}. Moreover, because the initial ignition is necessarily asymmetric, the ejecta of ash occurs preferentially along the initial ignition axis. 
Of this burned mass,  an even lesser fraction is ejected into the ISM, while  the rest of the ash mass remains gravitationally bound to the remnant star, and falls back on to the surface. Because of the asymmetric ejection of ash, the bound star gets kicked away from the ejecta with a speed of order $f_{\rm burned} v_{\rm ej}$, where $f_{\rm burned}$ is the fraction of the star that is burned, and $v_{\rm ej} \sim v_{\rm esc} \simeq 10^4$ km/s is a the ejection velocity, approximately the escape speed from a near-$M_{\rm Ch}$ WD. For  burn fractions $f_{\rm burn} \sim 0.1$, typical of SNe Iax simulations of CO WDs, the ejection velocity is $\sim 10^3$ km/s. This simple estimate is consistent with speeds of a few hundred to approximately one thousand km s$^{-1}$ observed in  hypervelocity  near-$M_{\rm Ch}$ SNe Iax primary WD remnants, classified as LP 40-365 objects. A total of six such candidate hypervelocity Iax remnants  are now known, lending support to this scenario \citep {raddietal18, elbadryetal23}.

Previous theoretical studies \citep {daveetal17, leungnomoto18} have focused upon models for the global nucleosynthetic abundance ratios of 3C 397. Spurred by the new ESA XMM-Newton abundance ratio measurements within localized clumps in 3C 397,  
in this paper, we will calculate synthetic clump abundance ratios. We will consider both DDT as well as pure deflagration   multidimensional hydrodynamical simulations whose global abundance ratios match 3C 397, and compute their resulting nucleosynthetic spatially-mapped yields. We will detail algorithms to isolate model clump analogs to the observed clumps. These model clumps will form the basis for comparison against 3C 397, and will address  key questions surrounding the SN Ia detonation mechanism and progenitor problems: What was the explosion mechanism that caused the SN Ia which gave rise to 3C 397: a DDT or a pure deflagration? How does the explosion mechanism impact the mixing of the ejecta, as seen in the XMM-Newton IGE abundances  spatially mapped  across the remnant? What are the implications for the broader SN Ia progenitor problem?

In the near-$M_{\rm Ch}$ scenario, the initial deflagration  shapes the ejecta, retaining the imprint of the Rayleigh-Taylor and Kelvin-Helmholtz instabilities (eg, \citealp{belletal04}). The subsequent evolution of the SNR also leads to the development of Rayleigh-Taylor instabilities at the interface between the remnant and the circumstellar medium \citep {mandaletal23}. Guided by the XMM-Newton observations of 3C 397, which clearly reveal anisotropies in the neutronized ejecta, here we focus upon the role of the initial deflagration in shaping the nucleosynthetic products of the explosion.

In \S 2, we lay out our methodology for the hydrodynamical simulations and nucleosynthetic post-processing of the ejecta. In \S 3, we present our results, including the algorithms employed for identifying model ejecta clumps to be compared directly against the observations of \citep {oshiroetal21}. In \S 4, we discuss our findings in the context of other observations and theoretical models. Finally, in \S 5, we conclude.

\section{Methodology} \label{sec:style}

\begin{table*}
\centering 
\resizebox{0.95\linewidth}{!}{
\begin{tabular}{llllll}
        \multicolumn{6}{c}{\textbf{Details of Models}} \\
        \toprule
Model Name   &  Explosion Mechanism  & $\mathrm{\rho_{c}(g\ cm^{-3})}$ & Bubble Offset\ (km) & Bubble Radius (km) & C/O Ratio \\
        \midrule
        DDT-HIGHDENS-HIGHOFF    & DDT                & $6.0 \times 10^{9}$ & 200 & 8  & 30/70 \\
        DEF-HIGHDENS-NR-CENTER  & Pure Deflagration  & $6.0 \times 10^{9}$ & 10  & 16 & 30/70 \\
        DEF-STDDENS-NR-CENTER   & Pure Deflagration  & $2.2 \times 10^{9}$ & 8   & 16 & 50/50
\end{tabular}}
\caption{The table above show the configurations of the hydrodynamical models that are presented in this paper.}\label{tab:Models}
\end{table*}

We utilize the \texttt{FLASH 4} \citep{fryxell_2000} hydrodynamical code with adaptive mesh refinement (AMR), which facilitates high resolution in the desired regions of the simulation domain. Our {\tt FLASH} simulations capture the nuclear burning within the flame which arises in the interior of near-$M_{\rm Ch}$ white dwarfs in 2D axisymmetric cylindrical geometry. These near-$M_{\rm Ch}$ models begin with a WD in hydrostatic equilibrium, parameterized by its central density and composition (C/O ratio and stellar progenitor metallicity). Because the initial near-$M_{\rm Ch}$ WD progenitor is convectively unstable due to carbon burning in the core, we generate WD initial conditions  using an adiabatic WD model generator \citep {chamulaketal08}. The WD models self-consistently include the simmering convective core with an adiabatic temperature gradient, using the same Helmholtz equation of state and carbon burning rate as is used in \texttt{FLASH} for self-consistency.  Based upon previous best fits to the global abundance ratios of Ni/Fe, Mn/Fe, and Cr/Fe in 3C 397 \citep {daveetal17}, we generate two WD progenitors for use in this study: one a high central density WD with central density $\rho_{c} =6.0 \times 10^{9}$ g cm$^{-3}$, and the other a standard central density WD $\rho_{c} = 2.2 \times 10^{9}$ g cm$^{-3}$. The initial composition of the high-central density ($\rm{6.0 \times 10^{9}}$ g cm$^{-3}$) WD model is 30/70 carbon-oxygen with a central temperature $T_{c} = 7.0 \times 10^{8}$ K. The standard central density ($\rm{2.2 \times 10^{9}}$ g cm$^{-3}$) WD model contains 50/50 carbon-oxygen to the initial composition with same T$_{c}$ of $7.0 \times 10^{8}$ K.  For our high central density WD model and standard central density WD model, we adopted the background temperatures of $\rm{10^{8}}$ K and $\rm{3 \times 10^{7}}$ K, respectively. The simulation begins with the mapping of the near-$M_{\rm Ch}$ mass WD model at the center of the simulation domain with initialization of hydrodynamic variables and passive Lagrangian tracer particles to be used for detailed nucleosynthetic postprocessing (see below) distributed by mass over the WD.

{\tt FLASH}   uses the Helmholtz equation of state to compute pressures and sound speeds, including electrons with an arbitrary level of degeneracy and relativity, radiation pressure, and non-degenerate nuclei. Local thermodynamic equilibrium between the gas and the radiation is assumed throughout, so that a single temperature applies to the electrons, positrons, ions, and radiation field. {\tt FLASH}  captures the nuclear-burning flame using a three-stage artifically-thickened flame burner,  modeled as  advection-diffusion reaction equations along with an unsplit higher-order Godunov hydrodynamics solver \citep {calderetal07}. 
The reactive hydrodynamics flame module contained within  {\tt FLASH} has energetics and neutronization calibrated against self-heating reaction networks, including neutronization, electron screening, and Coulomb corrections self-consistently \citep {calderetal07, townsleyetal16}. Three separate scalar fields solving separate advection diffusion reaction requations track carbon burning, burning to nuclear statistical quasi-equilibrium (NSQE), and burning to nuclear statistical equilibrium (NSE).

Our choice of hydrodynamical models is guided by our previous study, in which we matched the global abundance ratios of 3C 397 with two sets of models, one a high central density WD undergoing a DDT, and the second a pure deflagration of a standard central density WD \citep {daveetal17}. Table \ref{tab:Models} shows the initial conditions for the range of explosion models for the present study with the  mechanism, the WD central density, and the composition of the white dwarf including initial ignition bubble parameters such as ignition point offset from the center of the white dwarf, and initial bubble radius. This choice of models brackets the best fits to the global yields found in \citet {daveetal17}.

In considering the pure deflagration models, the question arises as to how to simultaneously select the ignition conditions while also varying the structure of the progenitor. 
Our strategy for this set of models is to keep the ignition parameters as relatively fixed, while varying the central density of the progenitor WD. We do so by scaling the offset radii to a characteristic length scale which arises when one compares the effects of buoyancy and deflagration \citep {fisherjumper15}. We determine a critical boundary for this offset radii using the equation \ref{eq:1}.
\begin{equation} \label{eq:1}
\mathrm{\displaystyle{\mathit{r}_{\mathrm{crit}} = 19\ \mathrm{km}\ \left (\frac{\mathit{S}_{l}}{100\ \mathrm{km}\ \mathrm{s}^{-1}} \right) \left (\frac{0.09}{\mathit{A}}\right)^{1/2} \left(\frac{2.2 \times 10^{9}\ \mathrm{g}\ \mathrm{cm}^{-3}}{\rho_{c}} \right)}}
\end{equation}
The critical offset radii is a function of laminar flame speed ($S_l$), the Atwood number $A  = (\rho_{f} - \rho_{a} ) / ( {\rho_{f} + \rho_{a}} )$, where $\rho_f$ and $\rho_a$ are the fuel and ash densities, respectively, and the central density $\mathrm{\rho_{c}}$. For ignition radii greater than this critical radius, $r_{\rm crit}$, the flame bubble rapidly ascends to the surface due to buoyancy. Conversely, when the offset distance falls below the critical offset, the flame bubble initially expands and consumes the dense core of the WD before buoyancy can take hold.
%
The critical radii are $\sim$16 km for the high central density model, and $\sim$14 km for the standard central density model. We choose a dimensionless offset radius, $r_0 / r_{\rm crit} = 0.6$ for the pure deflagration models, consistent with central ignitions. The corresponding physical offset radii $r_0$ are 10 km for the high central density WD and 8 km for the standard central density WD. Additionally, we also selected an initial bubble radius of 16 km in both pure deflagration models, and 8 km in the DDT model, all consistent with the parameters of \citet {daveetal17}. For our DDT model, we fix detonation initiation criterion at a  density of $\mathrm{2.6 \times 10^{7} g\ cm^{-3}}$; a detonation is triggered once the the flame front reaches this density.

The simulations are conducted with a 2D axisymmetric cylindrical geometry.  The $R=0$ boundary is axisymmetric by imposition, and diode boundary conditions are employed for the remaining three sides. The domain  spans 0 to $2^{17}$ km = $131072$ km along the cylindrical $R$-axis and -131072 km to 131072 km along $z$.  The maximum linear resolution resolution is 4 km at the highest refinement level 11, numbering the base level as 1. Our refinement criteria are jointly based upon both the density gradient and the flame progress variable. Regions of steep density gradients are resolved according to a standard \texttt{FLASH} criterion based upon the discrete dimensionless density gradient across cells \citep {lohner87, fryxelletal00}. Regions of active burning are always captured at the highest level of refinement \citep{townsleyetal07}. 
We fill the regions outside the WD with matter of low density with $\mathrm{10^{-3}\ g\ cm^{-3}}$ and a temperature of $\mathrm{3 \times 10^{7}\ K}$. This low density ``fluff"  has a negligible impact on the simulation other than to fill in the vacuum upon the Eulerian mesh outside the star. 
Modeling of 3C 397 suggests that the remnant is likely either in the later stages of free expansion or the earlier stages of the Sedov phase \citep {leahyranasinghe16}, based upon a distance range of  6.3 $\pm$ 0.1 - 9.7 $\pm$ 0.3 kpc, and an age range of 1350 - 1750 yr, which will justify comparisons  of the morphology obtained through our models versus that observed. 

Simulations incorporating a pure-deflagration explosion mechanism require an additional step to separate the ejected material from the bound WD that is kicked away from the ejecta. For this procedure, we employ a similar methodology as \citet {kashyapetal18}, using an energetic criterion to delineate bound and unbound material. We calculate the specific binding energy $e_{\rm b}$ for each particle at the last time step at the end of the simulation, $e_{\rm b} = e_{\rm int} +  \frac{1}{2}|v|^{2}+ \Phi$, where $e_{\rm int}$ is the specific internal energy, $\frac{1}{2} |v|^{2}$ the specific kinetic energy, and $\Phi$ the gravitational potential. Negative binding energy indicates the particle is gravitationally bound, and non-negative binding energy implies the particle is in the ejecta. While this energetic criterion provides a plausible estimate of the ejected mass, the general criterion for gravitationally unbound material in multidimensional hydrodynamics is not rigorously known, as hydrodynamical studies of common envelope evolution have shown \citep {ropkedemarco23}. In particular, we note that this criterion based upon the internal energy for the bound mass implicitly assumes that any additional mechanisms for mass loss, such as through thermally-driven or magnetically-driven winds, can be neglected. 

A further subtlety arises because the simulations do not possess a rigorously defined homologous expansion phase during the simulation time. This issue is particularly acute for pure deflagrations, which leave behind a bound remnant. Broadly speaking, in pure deflagrations, the outermost region in velocity space are ejected from the star, while the innermost ejecta will fall back onto the WD remnant. Near the surface of zero total energy, material is at the cusp of being unbound, and is sent out to large apocenter distances before either falling back or becoming marginally unbound. 
In the results section, we analyze the last few seconds of these simulations to assess the sensitivity of the ejecta to the cutoff time. 

The flame module suffices to capture the thermodynamics of the nuclear energy release accurately. The nuclear energy release is crucial for the hydrodynamical evolution,  but for detailed nucleosynthetic abundances needed to compare against SNR 3C 397, we must also utilize Lagrangian tracer particles. 10$^4$ constant mass tracer particles are embedded within the grid-based {\tt FLASH}  simulation and passively record the fluid density and temperature as a function of time. This number of particles typically establishes the nucleosynthetic yields within 10\% for a near-$M_{\rm Ch}$ SNe Ia model \citep {seitenzahletal10}. In post-processing, we first restructure time slices of particle data into Lagrangian thermodynamic temporal histories over individual particles. This post-processing step also includes a cut on the ejecta using the boundedness criterion identified above. In particular, gravitationally bound particles defined by this criterion will fall back on to the remnant, and are therefore not included in the ejecta yields included in this paper.

These Lagrangian density-temperature trajectories are post-processed with {\tt Torch}, using a 489 isotope nuclear network \citep{timmes1999integration}. In contrast, the flame model employed  inline in {\tt FLASH} tracks the effective electron molar fraction $Y_{e}$ along with the flame progress variables, but does not track detailed abundances \citep {calderetal07}. The estimated 1350-1750 yr age of 3C 397  \citep{leahyranasinghe16} is sufficient for  most of the unstable iron group species to decay into stable species. However, the radioisotope $^{53}$Mn subsequently decays into stable $^{53}$Cr by  electron capture with a 3.7 Myr half life. Similarly, the radioactive decay of  $^{59}$Ni to $^{59}$Co occurs primarily through electron capture with a $7.6 \times 10^4$ kyr half life. 
In our study, we include the parent nuclei $^{53}$Mn and $^{59}$Ni as effectively stable over the lifetime of 3C 397. Throughout this paper, all quoted model atomic abundances are derived from these modified decayed stable isotopes.

The ratios of the decayed stable atomic abundances obtained from our model pipeline can then be directly compared against the  ratios measured for 3C 397.  To understand the effect of metallicity, we also varied the initial metallicity by including $\mathrm{^{22}Ne}$ as a proxy for the WD stellar progenitor metallicity, as the result of reaction bottlenecks during the CNO cycle in  WD progenitors \citep {timmesbrowntruran03}. We considered same solar metallicity ($Z_{\odot}$) from \citep{asplundetal09} as previously used by \citep{daveetal17}. Studies of the neutronization during the simmering phase leading up to unstable deflagration ignition have generally found that the effect of simmering is less significant (with neutron excesses of order $10^{-4}$) \citep {pirobildsten08, martinezrodriguezetal16}.
These post-processed yields are aggregated into  spatially mapped abundances. {\it The spatially-mapped abundances from the 2D simulations are crucial to understanding the asymmetry evident in the recently resolved XMM-Newton data of 3C 397.}


\section {Results} \label {sec:results}

\begin{figure*}

    \begin{subfigure}{0.80\textwidth}
        \centering
        \caption{DDT-HIGHDENS-HIGHOFF}
        \label{fig:ejecta_im1}
        \includegraphics[width = 0.80\linewidth]
        {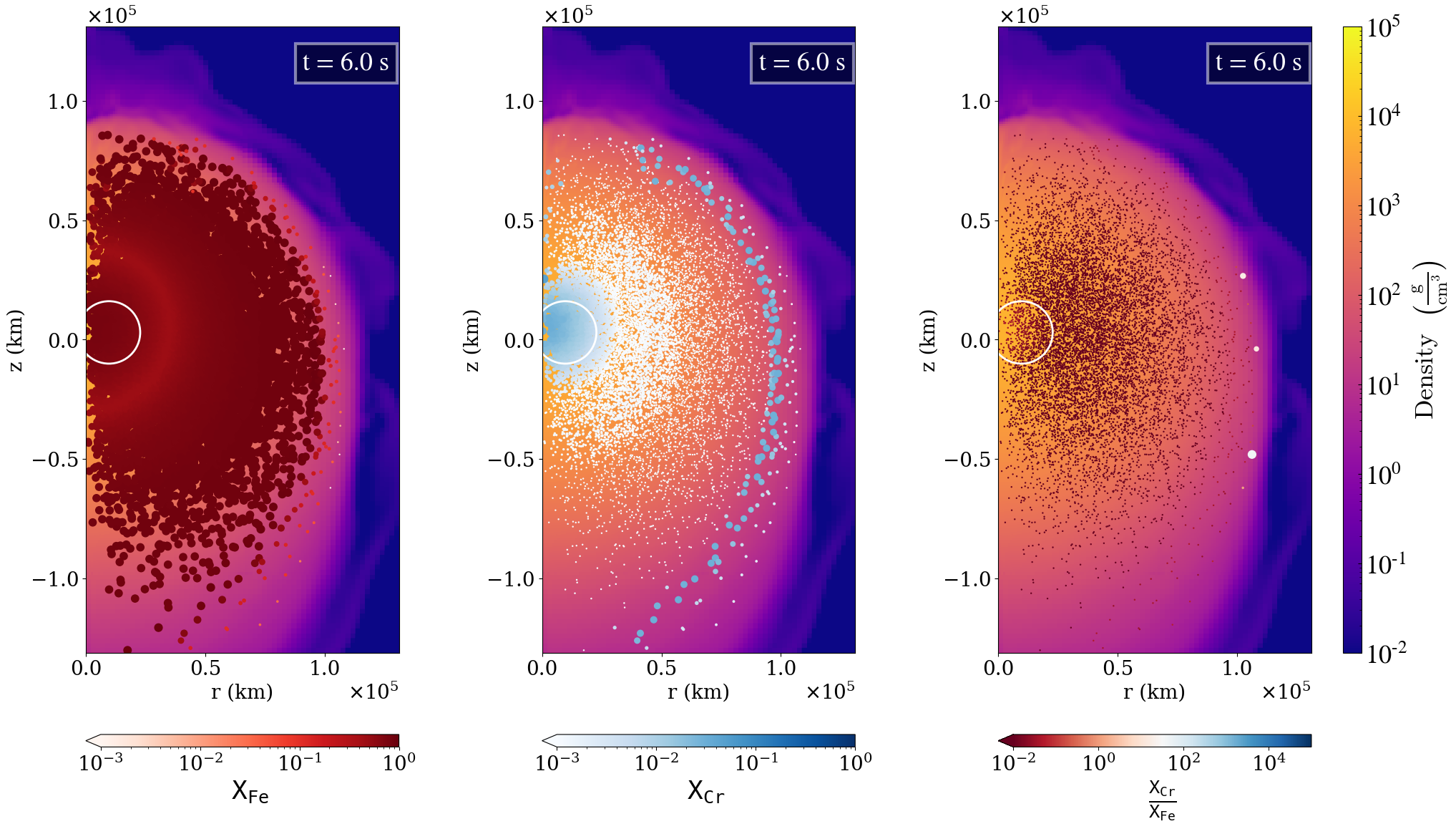}
    \end{subfigure}
       
    \begin{subfigure}{0.80\textwidth}
        \centering
        \caption{DEF-HIGHDENS-NR-CENTER}
        \label{fig:ejecta_im2}
        \includegraphics[width = 0.80\linewidth]
        {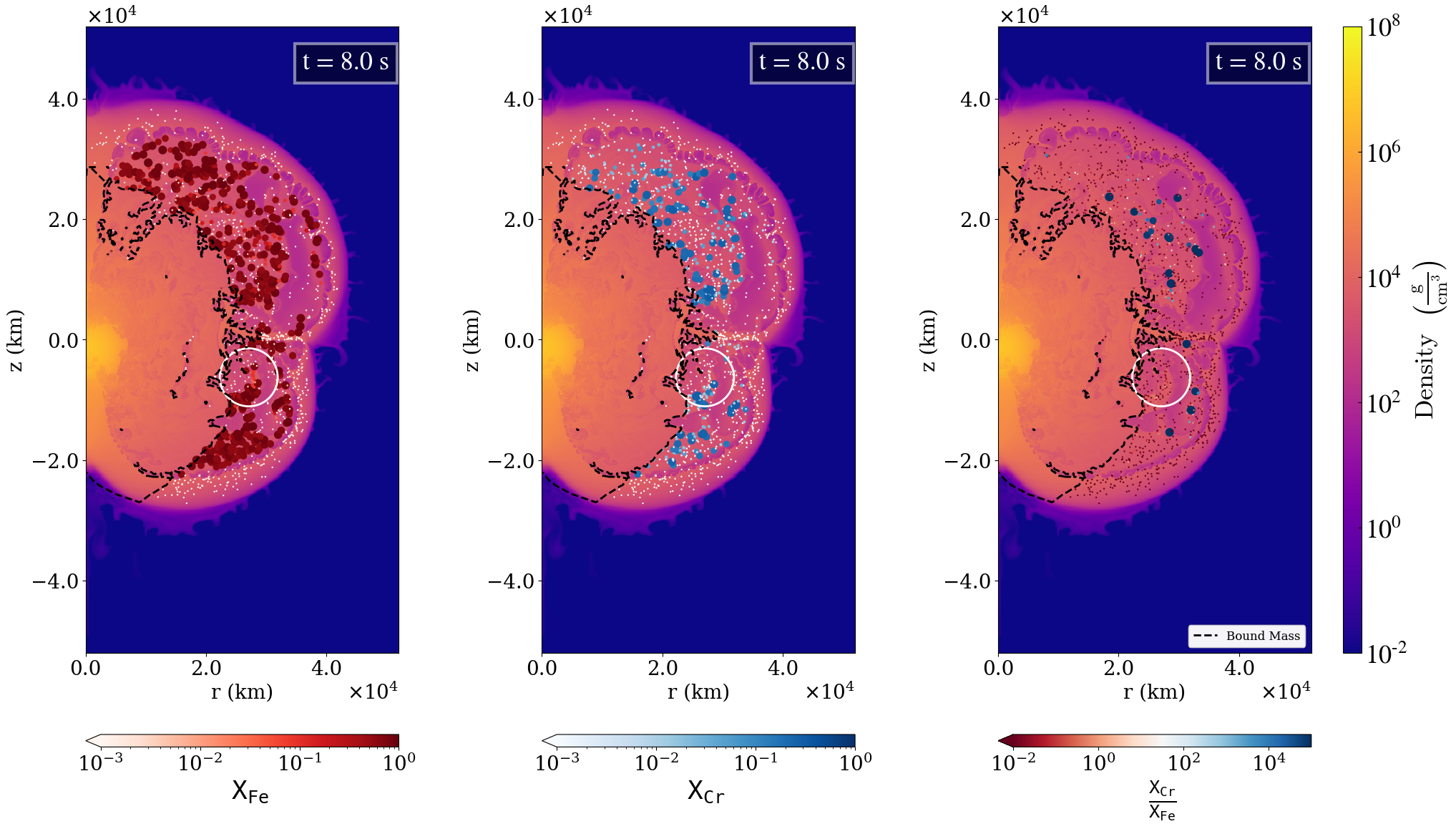}
    \end{subfigure}

    \begin{subfigure}{0.80\textwidth}
        \centering
        \caption{DEF-STDDENS-NR-CENTER}
        \label{fig:ejecta_im3}
        \includegraphics[width = 0.80\linewidth]
        {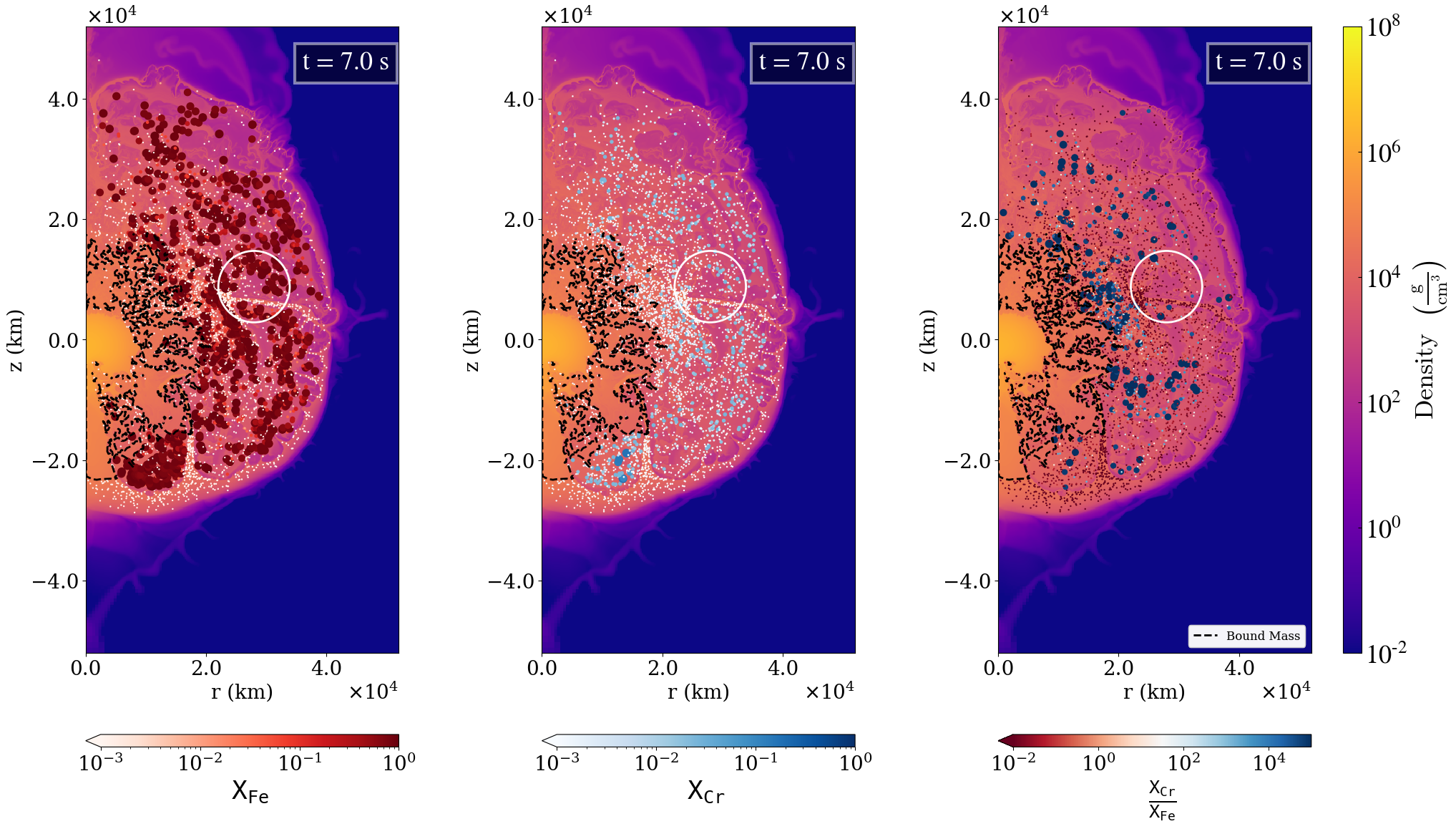}
    \end{subfigure}

\centering
\caption{Plots comparing the outcome of the simulations at end-time. Shown, top-to-bottom, are models (a) DDT-HIGHDENS-HIGHOFF, (b) DEF-HIGHDENS-NR-CENTER, and (c) DEF-STDDENS-NR-CENTER. Note that the pure deflagration models have lower expansion velocities, and so are shown on smaller domains than the DDT model. The right-side color bar represents the log of the mass density. Additionally, each figure shows the log Fe abundance $X_{\rm Fe}$, the log Cr abundance $X_{\rm Cr}$, and the log Cr/Fe abundance ratio $X_{\rm Cr} / X_{\rm Fe}$ on unbound tracer particles. In order to highlight the clumps, the size of the tracers is scaled proportional to the abundance and abundance ratios. All models depicted have metallicity $Z = 3.0 Z_{\odot}$. The annotated black dashed line contours in pure-deflagration models enclose the bound mass. The annotated circles show the highest likelihood clumps that are identified for all models; see Section \ref{subsec: Clump Identification and Analysis}.
}
\label{fig:spatial_distribution}
\end{figure*}

\begin{figure*}
    \begin{subfigure}{0.33\textwidth}
        \centering
        \includegraphics[width=0.95\linewidth]{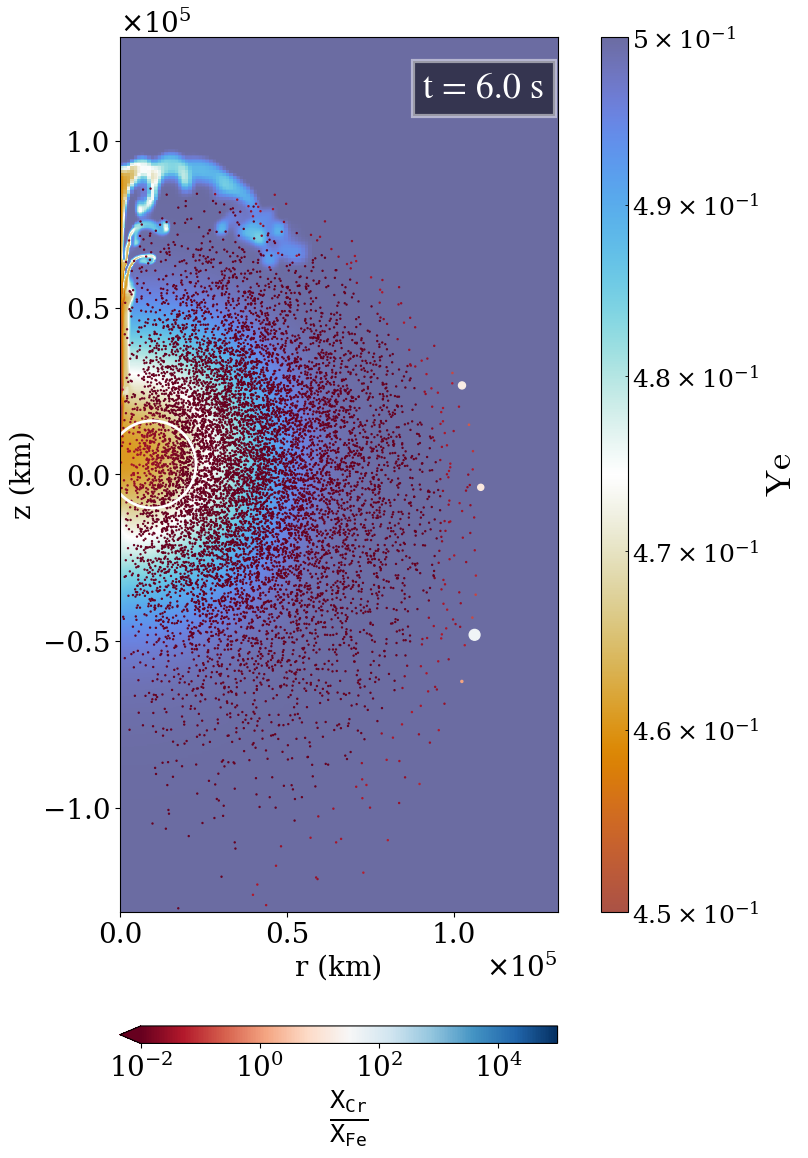} 
        \caption{DDT-HIGHDENS-HIGHOFF}
        \label{fig:subim1}
    \end{subfigure}
    \begin{subfigure}{0.33\textwidth}
        \centering
        \includegraphics[width=0.95\linewidth]{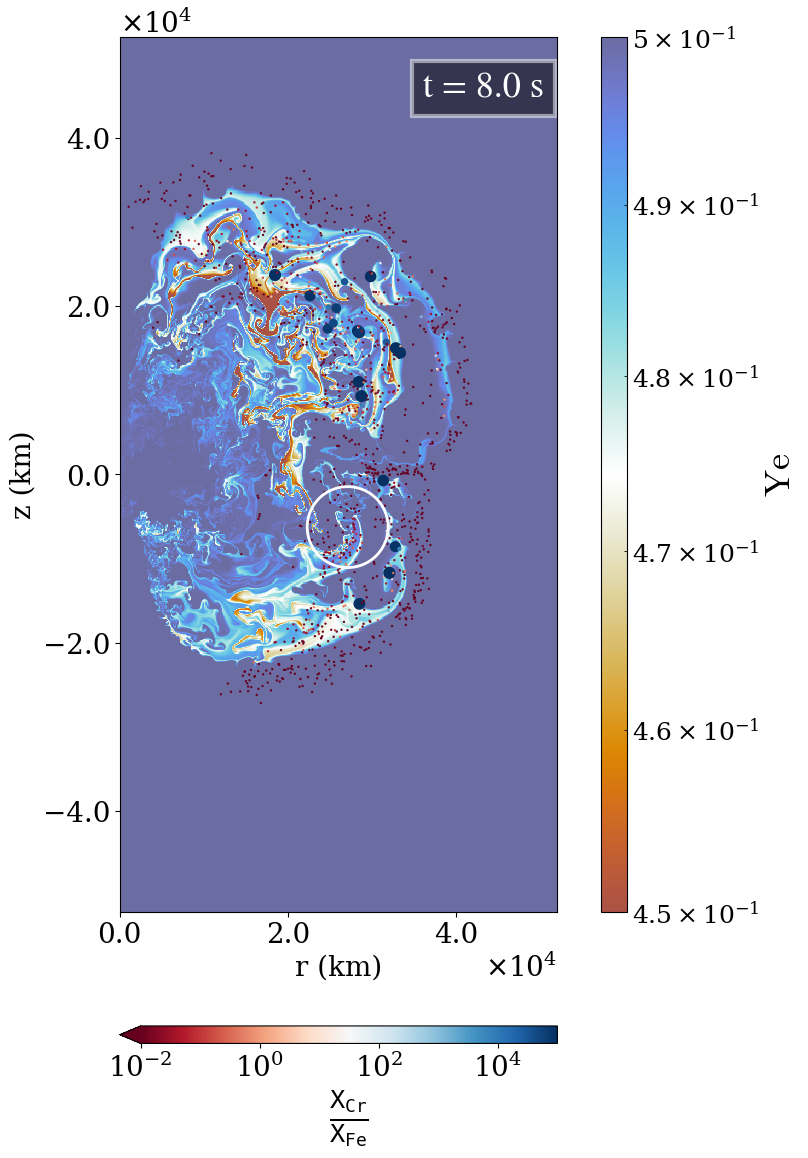}
        \caption{DEF-HIGHDENS-NR-CENTER}
        \label{fig:subim2}
    \end{subfigure}
    \begin{subfigure}{0.33\textwidth}
        \centering
        \includegraphics[width=0.95\linewidth]{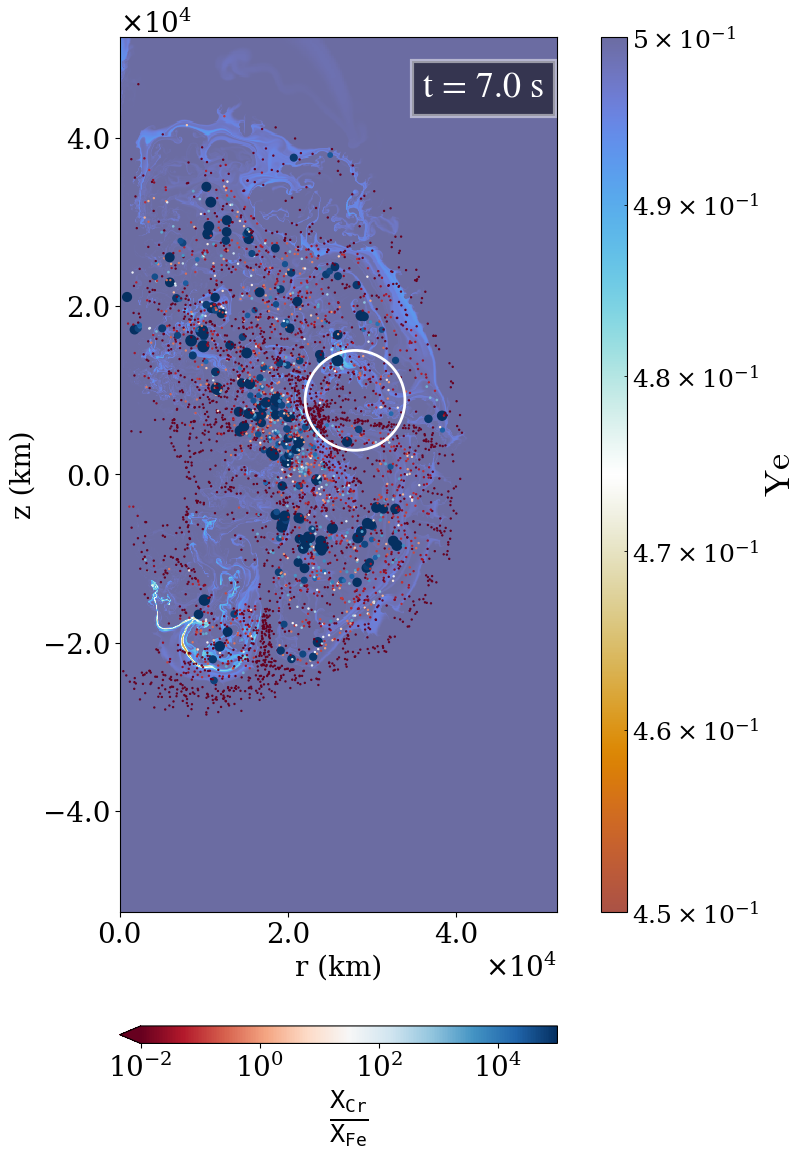}
        \caption{DEF-STDDENS-NR-CENTER}
        \label{fig:subim3}
    \end{subfigure}

    \caption{Plots comparing the distribution of electrons per baryons (Y$_{e}$) among our three models : (a) DDT-HIGHDENS-HIGHOFF, (b) DEF-HIGHDENS-NR-CENTER, and (c) DEF-STDHDENS-NR-CENTER.  
    Over-plotted particles represents Cr/Fe abundance ratio on log scale. The annotated circles show groups of particles that contains 7\% of the total iron in the ejecta with the highest nucleosynthetic likelihood to match the southern 3C 397 clump. All models depicted have metallicity $Z = 3.0 Z_{\odot}$.}
    \label{fig:clump_image}
\end{figure*}

\subsection{Spatial Distribution of the Neutronized Species}

Figure \ref{fig:spatial_distribution} shows density slice plots of our three models. In particular, figure \ref{fig:ejecta_im1} showcases DDT-HIGHDENS-HIGHOFF model at the time t = 6s, figure \ref{fig:ejecta_im2} displays DEF-HIGHDENS-NR-CENTER model at the time t = 8s, and figure \ref{fig:ejecta_im3} illustrates  the DEF-STDDENS-NR-CENTER model at the time t = 7s. The color bar on the right side of the figures indicates density on the log scale.  The overplotted distribution of particles in the figures shows the abundance of Fe, Cr, and the abundance ratio of Cr/Fe in the ejecta. 
The left column shows  Fe abundances, the center column Cr abundances, and the right column the abundances ratios of Cr/Fe. For clarity, the particle size is scaled proportional to the mass fraction or mass fraction ratio. The color bar below each column shows the range of mass fractions on a log scale, with iron-rich tracer particles in red, and chromium-rich tracers  in blue. 

These plots unveil  
the underlying connections between the ejecta morphology, the explosion mechanisms, and the distribution of neutron-rich elements in each model. Because the supersonic detonation propagates faster than the hydrodynamical response time of the WD, the density profiles of the DDT-HIGHDENS-HIGHOFF model reveal the hallmark of the DDT: a smoothly-stratified structure of the burned material in the ejecta. This feature is highlighted further in Figure \ref{fig:clump_image}, which overlays the tracer particles shaded by Cr/Fe as in Figure \ref{fig:spatial_distribution}, but now shown overplotted against the electron fraction $Y_e$, defined as $Y_e = \sum Z_i X_i / A_i$, where $Z_i$ and $A_i$ are the atomic and mass numbers of isotope $i$ with mass fraction $X_i$. The flame module used in FLASH tracks $Y_e$ on the Eulerian mesh, and it is this field which is plotted in Figure  \ref  {fig:clump_image}. In particular, the abundances of neutronized regions (where $Y_e < 0.5$) for model DDT-HIGHDENS-HIGHOFF are concentrated in the central region with a lesser amount along the outskirts, the result of ash ejected from the WD just prior to detonation. The clump identified as having the highest mean Cr/Fe ratio is nearly centrally located for this model.
Furthermore, we also confirm the released kinetic energy in this DDT simulation is  $1.43 \times 10^{51}$ erg, with the outer ejecta moving at approximately 10,000 km s$^{-1}$. We calculate the mass of $^{56}$Ni in the ejecta of DDT-HIGHDENS-HIGHOFF to be 0.76 M$_{\odot}$, which would roughly correspond to a bright normal SN Ia in terms of its expected light curve.

In marked contrast, the failed detonation models DEF-HIGHDENS-NR-CENTER and DEF-STDDENS-NR-CENTER, shown in (b) and (c) on Figures  \ref {fig:spatial_distribution} and \ref  {fig:clump_image}  exhibit highly asymmetric ejecta, both in terms of their mass as well as their neutronized isotope distributions. Critically, unlike the smooth structure of neutronization within the DDT model, the distribution of neutronized isotopes in the ejecta of DEF-HIGHDENS-NR-CENTER and DEF-STDDENS-NR-CENTER are highly aspherical and clumpy. {\it Fundamentally, this is because the pure deflagration models never transition into detonation, causing the neutronized ash buoyed up from the dense core to maintain the complex morphological structures developed during the onset of the Rayleigh-Taylor and Kelvin-Helmholtz instabilities during burning} \citep {woosleywunschkulhen04, belletal04, hicks15}. These structures are frozen in as the ejecta enter into homologous expansion, where they contribute to the diversity of the class of SNe Iax \citep {lachetal22}. 

The released kinetic energy in both DEF-HIGHDENS-NR-CENTER and DEF-STDDENS-NR-CENTER models are $\sim 8 \times 10^{49}$ erg and $\sim 1.7 \times 10^{50}$, and the ejecta velocity as large as a few thousand km s$^{-1}$. Our analysis confirms that large portions of the WD masses are still bound at the end of these two simulations. In particular, the DEF-HIGHDENS-NR-CENTER model ejects about 0.19 $\mathrm{M_{\odot}}$ and DEF-STDDENS-NR-CENTER model ejects about 0.56 $\mathrm{M_{\odot}}$, leaving behind remnants of 1.2 $\mathrm{M_{\odot}}$ and 0.82 $\mathrm{M_{\odot}}$, respectively. Similar to previous investigations of pure deflagrations, we find the $^{56}$Ni ejected by the HIGHDENS-NR-CENTER and DEF-STDDENS-NR-CENTER models to be an order of magnitude or less than produced in a normal SN Ia event. HIGHDENS-NR-CENTER and DEF-STDDENS-NR-CENTER nucleosynthesize $\sim$ 0.032 $\mathrm{M_{\odot}}$ and $\sim$ 0.084 $\mathrm{M_{\odot}}$ of $^{56}$Ni ejecta, respectively.


\subsection{Clump Identification and Analysis} \label{subsec: Clump Identification and Analysis}


A key goal of this study is to compare the clumps formed in simulations of SNR 3C 397 directly against the recent XMM-Newton observations of \citet {oshiroetal21}. Accordingly, we proceed to constrain the model clumps rich in iron group elements Ni, Mn, Cr, and Fe, and the newly-discovered Ti,  all identified in X-ray spectra of two clump regions of 3C 397 (``south" and ``west").

The extraordinarily high abundances of Ti and iron group elements observed in the southern clump of 3C 397  are consistent with burning in dense environments  \citep{oshiroetal21, leungnomoto18}.  Fundamentally, this is because efficient electron captures occur during burning at high densities, reducing the number of free electrons, and shifting nucleosynthesis yields towards neutron-rich abundances ($Y_e < 0.5$) \citep{Lach_2020, Mori2020}. 
As discussed in \citet {daveetal17}, the resulting Cr/Fe abundance ratios are particularly sensitive to the density of the burning. Because the southern clump has a higher Cr/Fe abundance, it poses the tightest constraints on the burning conditions, and we focus upon it solely here.

We identify clusters of particles based on nucleosynthetic abundances and spatial distribution. To analyze model clumps, our algorithm proceeds via an exhaustive search as follows. We loop over all $N_{\rm ej}$ ejected tracer particles, picking each particle one at a time to represent the center of a localized candidate clump. We subsequently add up the Fe from its nearest neighbor particles, selected using the kd-tree method \texttt{scipy.spatial.KDTree} \citep {friedmanetal77}, until the cumulative mass of Fe in the candidate clump reaches the same 7\% of the overall iron abundance in the model ejecta as that estimated in the southern clump of 3C 397 by \citet{oshiroetal21}. Because we fix the total Fe mass in each candidate clump, the total mass or equivalently the number of particles will vary both between clumps and between models.
This process of determining  the $N_{\rm ej}$ candidate clumps is repeated in turn for each of the $N_{\rm ej}$ ejecta particles in the dataset. Because the pure deflagration models lead to lower ejecta masses, the total mass of iron in the ejecta and also the cumulative iron mass in the particle clumps also varies. We note that in the DDT-HIGHDENS-HIGHOFF models, the total iron enclosed by each candidate clumps is $\sim$ 0.07 M$_{\odot}$, while the clumps in DEF-HIGHDENS-NR-CENTER models comprise $\sim$ 0.0034 M$_{\odot}$ of iron. The DEF-STDDENS-NR-CENTER models shows $\sim$ 0.0067 M$_{\odot}$ of iron in the clumps.

In our next step, we compared the nucleosynthetic yields of candidate particle clusters in our simulations to the southern clump observed in SNR 3C 397. We define a likelihood function 
that compares the abundance ratios from our models to the observations:
\begin{equation} \label{eq:2}
\ln(\text{likelihood}) = -\frac{1}{2} \sum_{i} \left (\frac{(X_{\text{obs},i} - X_{\text{model},i})^{2}}{\sigma_{i}^{2}} + \ln (2 \pi \sigma_{i}^{2})\right)
\end{equation}
The index $i$ runs over the abundance ratios of Cr/Fe, Mn/Fe and Ni/Fe. $X_{\text{obs},i}$ and $X_{\text{model},i}$ are the abundance ratios of the observed clump and model clusters, respectively. ${\sigma_i}$ are the observational abundance ratio errors. 

We also examine the model statistical errors associated with  Cr/Fe, Ni/Fe, Mn/Fe, and Ti/Fe within the entire mass associated with the clump. We utilize the jackknife resampling method \texttt{astropy.stats.jackknife\_stats} \citep{quenouille56} to elegantly capture the statistical uncertainty associated with our clump analysis procedure. The jackknife method systematically resamples the clump, removing one particle at a time, and estimates from these resamples the confidence interval of individual isotopic abundances and mass ratios for the cluster. The clumps each have of order $10^2$ particles, so the discrete error associated with resampling produces uncertainties are typically on the a few percent level. These model statistical errors are much less than the observational error bars, which are of the order of tens of percent. Consequently, the inclusion of only the experimental errors in the likelihood function in Equation \ref {eq:2} is self-consistent.

Finally, we isolate the single most likely clump for each simulation model with the highest likelihood values, out of all $N_{\rm ej}$ candidates, and compare directly against the southern clump identified by \citet {oshiroetal21}  in 3C 397. This final selection is then the  most likely clump with an equivalent Fe mass obtained by \citet {oshiroetal21}. We quoted the abundance ratios from these most likely clumps in each models (including different initial progenitor metallicity) in Table \ref{tab:Mass_fractions} along with the statistical confidence intervals obtained from the jackknife statistical analysis. We highlighted the most likely clump for each model by a white circle in Figures \ref{fig:spatial_distribution} and \ref{fig:clump_image}. 

    \begin{table*}
    \centering
    \resizebox{0.8\linewidth}{!}
    {%
    \begin{tabular}{c|cccc}
    \multicolumn{5}{c}{\textbf{DDT-HIGHDENS-HIGHOFF}} \\
    \hline
    \multicolumn{1}{c|}{Metallicity (Z)} & 
    \multicolumn{1}{c}{Ni/Fe}  & 
    \multicolumn{1}{c}{Mn/Fe} & 
    \multicolumn{1}{c}{Cr/Fe} & 
    \multicolumn{1}{c}{Ti/Fe} \\
    \hline 
               
                          
                         
    $Z = 0$  & $ 0.34947^{+0.00316}_{-0.00319}$ &
               $ 0.04170^{+0.00067}_{-0.00068}$ &
               $ 0.01652^{+0.00059}_{-0.00060}$ & 
               $ (1.6\times10^{-6})^{+7.9\times10^{-8}}_{-8.0\times10^{-8}} $ \\
               
    $Z = 1.5 Z_{\odot}$ & $ 0.33568^{+0.00295}_{-0.00298}$ &
                          $ 0.04018^{+0.00073}_{-0.00074}$ &
                          $ 0.01741^{+0.00060}_{-0.00056}$ &
                          $ (1.7\times10^{-6})^{+9.3\times10^{-8}}_{-9.4\times10^{-8}} $ \\
                          
    $Z = 3.0 Z_{\odot}$ & $ 0.31815^{+0.00260}_{-0.00262}$ &
                          $ 0.03804^{+0.00079}_{-0.00079}$ &
                          $ 0.01865^{+0.00061}_{-0.00062}$ &
                          $ (1.9\times10^{-6})^{+1.1\times10^{-7}}_{-1.1\times10^{-7}} $ \\[3pt]
    
    \hline 
    \multicolumn{5}{c}{} \\
    \multicolumn{5}{c}{\textbf{DEF-HIGHDENS-NR-CENTER}} \\
    \hline
    \multicolumn{1}{c|}{Metallicity (Z)} & 
    \multicolumn{1}{c}{Ni/Fe}  & 
    \multicolumn{1}{c}{Mn/Fe} & 
    \multicolumn{1}{c}{Cr/Fe} & 
    \multicolumn{1}{c}{Ti/Fe} \\
    \hline
               
    
    $Z = 0$  & $ 0.30450^{+0.01286}_{-0.02077} $ &
               $ 0.03686^{+0.00121}_{-0.00195} $ &
               $ 0.13226^{+0.02053}_{-0.03316} $ & 
               $ 0.02641^{+0.01153}_{-0.01862} $ \\
               
    $Z = 1.5 Z_{\odot}$ & $ 0.21491^{+0.01789}_{-0.03143} $ &
                          $ 0.03298^{+0.00046}_{-0.00081} $ &
                          $ 0.08887^{+0.01676}_{-0.02944} $ &
                          $ 0.00846^{+0.00440}_{-0.00773} $ \\
                          
    $Z = 3.0 Z_{\odot}$ & $ 0.20101^{+0.01729}_{-0.02992} $ &
                          $ 0.03086^{+0.00045}_{-0.00078} $ &
                          $ 0.08787^{+0.02246}_{-0.03888} $ &
                          $ 0.01212^{+0.00497}_{-0.00860} $ \\[3pt]
    \hline
    \multicolumn{5}{c}{} \\
    \multicolumn{5}{c}{\textbf{DEF-STDDENS-NR-CENTER}} \\
    \hline
    \multicolumn{1}{c|}{Metallicity (Z)} & 
    \multicolumn{1}{c}{Ni/Fe}  & 
    \multicolumn{1}{c}{Mn/Fe} & 
    \multicolumn{1}{c}{Cr/Fe} & 
    \multicolumn{1}{c}{Ti/Fe} \\
    \hline
               
    
    $Z = 0$  & $ 0.12422^{+0.00990}_{-0.01582} $ &
               $ 0.01315^{+0.00105}_{-0.00167} $ &
               $ 0.01278^{+0.00122}_{-0.00195} $ & 
               $ 0.00056^{+5.6\times10^{-5}}_{-9.0\times10^{-5}} $ \\
               
    $Z = 1.5 Z_{\odot}$ & $ 0.14805^{+0.01480}_{-0.02397} $ &
                          $ 0.01780^{+0.00049}_{-0.00080} $ &
                          $ 0.01199^{+0.00105}_{-0.00170} $ &
                          $ 0.00047^{+4.9\times10^{-5}}_{-7.9\times10^{-5}} $ \\
                          
    $Z = 3.0 Z_{\odot}$ & $ 0.19729^{+0.01417}_{-0.02220} $ &
                          $ 0.02764^{+6.7\times10^{-5}}_{-1.0\times10^{-4}} $ &
                          $ 0.01247^{+0.00085}_{-0.00134} $ &
                          $ 0.00043^{+4.5\times10^{-5}}_{-7.0\times10^{-5}} $ \\[3pt]
    \hline
    \multicolumn{5}{c}{} \\
    \multicolumn{5}{c}{\textbf{Observed clumps in the SNR 3C 397}} \\
    \hline
        \multicolumn{1}{c|}{Region} & \multicolumn{1}{c}{Ni/Fe}  & \multicolumn{1}{c}{Mn/Fe} & \multicolumn{1}{c}{Cr/Fe} & \multicolumn{1}{c}{Ti/Fe} \\
    \hline
    South & $ 0.18^{+0.07}_{-0.03} $ &
        $ 0.051^{+0.009}_{-0.009} $ &
        $ 0.106^{+0.011}_{-0.009} $ &
        $ 0.014^{+0.004}_{-0.004} $ \\
    
    West  & $ 0.27^{+0.05}_{-0.03} $ & 
        $ 0.050^{+0.006}_{-0.006} $ & 
        $ 0.034^{+0.004}_{-0.004} $ & 
        $ < 0.002 $ \\ [3pt]
    \hline
                          
    \end{tabular}
    }
\caption{The tables above shows the localized clumps' abundance ratios of Ni/Fe, Mn/Fe, Cr/Fe and Ti/Fe from DDT-HIGHDENS-HIGHOFF, DEF-HIGHDENS-NR-CENTER, and DEF-STDDENS-NR-CENTER models. The chosen clumps have the highest nucleosynthetic likelihood and contains  7\% of the total iron mass in the ejecta. Three rows in each section indicates the different metallicities $Z = 0$, $Z = 1.5 Z_{\odot}$ and $Z = 3.0 Z_{\odot}$ of the initial WD progenitor model. The solar metallicity ($Z_{\odot}$) is taken from \citep{asplundetal09}. For comparison the mass fraction ratios of observed clump in the southern and western regions of SNR are also reported.} 
\label{tab:Mass_fractions}
    \end{table*}

Inspecting the most likely clumps from each of the hydrodynamic models, we determined the most likely particle cluster in the DDT-HIGHDENS-HIGHOFF model (Figure \ref{fig:ejecta_im1}) is situated at the innermost parts of the systematically layered ejecta. However, the Cr/Fe abundance ratio of this particle cluster is less than one-fifth of the reported value from the observed clump. This local clump analysis establishes that a DDT even within a high central density WD does not achieve the nucleosynthesis observed within the local neutronized clumps of 3C 397.  Furthermore, the centralized location of the model clump in the inner ejecta  disagrees with the observed clump's outer location in SNR 3C 397.

An identical comparison in DEF-STDDENS-NR-CENTER and the DEF-HIGHDENS-NR-CENTER models (Figures \ref{fig:ejecta_im3} and \ref{fig:ejecta_im2}) shows the most likely particle cluster in the pure deflagrations are centered towards the outskirts of the ejecta. The Cr/Fe ratio of the most likely cluster from the  DEF-STDDENS-NR-CENTER model is nearly one-eighth of the value reported to clump's observation data. In contrast, the DEF-HIGHDENS-NR-CENTER model show multiple particle groups in the ejecta that have Cr/Fe ratios comparable to, and in some cases even higher than, the Cr/Fe ratios reported in the observations. 
The most likely DEF-HIGHDENS-NR-CENTER model cluster has Cr/Fe abundance ratio more closely aligned with the corresponding observational value. 

Our analysis shows that overall, the most likely model with respect to both morphology and nucleosynthesis is the high central density model with a pure-deflagration explosion mechanism (DEF-HIGHDENS-NR-CENTER) and initial metallicity of 3.0 Z$_{\odot}$. For this model, the clump Mn/Fe and Cr/Fe ratios are within 1 and 3 $\sigma$ of the observed values, respectively. The Ni/Fe ratios are within 3 $\sigma$. The Ti/Fe is in tension with the observations at the 4 $\sigma$ level. 
In contrast, both the DDT-HIGHDENS-HIGHOFF and DEF-STDDENS-NR-CENTER models are in large disagreement with observation, notably with Cr/Fe disagreements at 9 $\sigma$ or greater.
Overall, bearing in mind that we have not fine tuned our hydrodynamic models to match the observations, the model clump nucleosynthesis and morphology overwhelmingly favor the DEF-HIGHDENS-NR-CENTER model.

\subsection {Sensitivity of Ejecta Criterion}
As mentioned in the methods section, we analyze the increment in the unbound ejecta mass from the marginally bound ejecta over fixed intervals in time in order to assess the sensitivity of the ejecta to the time cutoff of the simulation. In DEF-HIGHDENS-NR-CENTER simulation, the ejecta mass increment between 5s and 6s is $\sim$ 12\%, between 6s and 7s is $\sim$ 7\%, and between 7s and 8s is $\sim$ 5.3\%. In DEF-STDDENS-NR-CENTER simulation, the increment between 4s and 5s is $\sim$ 85.5\%, between 5s and 6s is 36.6\%, and between 6s and 7s is $\sim$ 14.2\%. 

Nonetheless, once the expelled material becomes gravitationally unbound from the star, the radial distribution of the mass does not alter significantly under a free expansion. 
To quantify the degree to which to unbound ejecta achieves free expansion, we follow an analysis procedure reported in \citet{roepke05}. We divide the ejecta particles into radial shells of equal width and find the mean velocity of particles in each shell. We then calculate the deviations from the radial velocities from a linear homologous expansion. In our DEF-HIGHDENS-NR-CENTER model, the maximum deviation is 17\%, while the mass weighted deviation remains less than 3\%.  The weighted deviation in the DEF-STDDENS-NR-CENTER model also remains less than 3\%. These results confirm that that the ejecta criteria robustly identifies the ejecta material in homologous expansion, even in models with bound remnants.

\section {Discussion} \label {sec:disc}

As previous authors have established, buoyancy-driven plumes generally lead to the ejection of neutronized material to the outer edge of the ejecta even in the DDT scenario \citep [e.g.,][]{seitenzahletal13b}. We note that our DDT simulation also has a plume of neutronized ejecta along the +$z$ axis, where the initial deflagration ignition point was located (Figure \ref {fig:clump_image}a).  We are agnostic with regards to where the selected model clump originates; in principle this outermost region of DDT ejecta could have been selected if it had been the best fit to the 3C 397 southern clump nucleosynthetic abundances. However, for the DDT model, it is the central region and not the outer one that produces a greater neutronization and a better match to the southern clump data. Figure \ref {fig:clump_image}a demonstrates  that the buoyancy-driven deflagration ash is less neutronized than the material in the core, as it is burned at lower densities on average. Thus the buoyancy-driven effect in the DDT model considered here is not sufficient to explain the neutronization in the clumps observed in 3C 397. 



Thermal broadening in explosive transients is significantly larger than the isotopic shifts for nuclei, limiting optical and X-ray observations of SNe and their remnants to elemental rather than isotopic abundances. With the exception of gamma-ray spectroscopy and monoisotopic elements like $^{55}$Mn, astronomical observations generally cannot probe the nuclear isotopes crucial for nuclear astrophysics.
In a nutshell, with the exception of gamma-ray spectroscopy and monisotopes like $^{55}$Mn, astronomical observations do not probe the nuclear isotopes so critical for nuclear astrophysics. In contrast, the isotopic abundances of meteorites can be analyzed in detail in the laboratory. The clumpy structures prevalent in the pure deflagration models support the idea that exceptionally high abundances of neutronized isotopes may persist and be found in meteoretic clumps \citep {oshiroetal21}. The highly neutronized isotopes formed in 3C 397 point towards supersolar abundances of IGE and IMGE isotopes such as $^{54}$Cr, $^{50}$Ti, $^{58}$Fe, $^{62}$Ni, and $^{48}$Ca. The carbonaceous chondrite meteors Murchison and Allende have long been known to have inclusions exhibiting correlated overabundance in both $^{48}$Ca and $^{50}$Ti, with large variations in inferred neutron excess across grains \citep {zinneretal86}. More recent measurements of the Orgueil CI meteorite have also revealed high abundances of $^{54}$Cr and $^{50}$Ti \citep {nittleretal18}. These measurements excessively abundant in neutronized isotopes are consistent with both nucleosynthesis in electron capture SNe (ECSNe) as well as in high-density WD SNe Ia progenitors \citep {nittleretal18}. As we have found, the neutronized material inside a DDT is concentrated towards the center of the remnant, while the dense WD SNe Iax progenitors naturally produce clumpy distributions of neutronized material in the outer ejecta, lending support to the idea that these clumps may survive and become incorporated into meteoretic grains. Moreover, a direct comparison against the localized clumps such as those presented here, versus the global nucleosynthetic yields will be a more faithful representation of the actual composition of the neutronized grains. 


In the pure deflagration scenario for the origin of SNe Iax, the asymmetric ejection of the ash imparts a kick to the primary WD, causing it to be ejected from the system as a hypervelocity WD \citep {jordanetal12b}. Six hypervelocity candidates have now been discovered thanks to the Gaia survey \citep {elbadryetal23}. Based on our finding that 3C 397 likely originated from a pure deflagration SN Ia, the single degenerate scenario for near-$M_{\rm Ch}$ WDs predicts that both the stripped non-degenerate donor as well as the kicked accretor should have survived the explosion. The 8 - 9.7 kpc estimated distance \citep {leahyranasinghe16} to 3C 397 makes a direct search for both survivors implausible through Gaia. 

Qualitatively, the asymmetric structure of the ejecta introduced by the pure-deflagration scenario is consistent with the extended Fe-K shell emission in 3C 397 \citep {oshiroetal21}.
On the other hand, the low explosion energy and low mass ejecta inferred in the presented study seem to be in tension with the bright luminosity of the Fe-K emission \citep {yamaguchietal14}. 
In order to clarify whether the characteristics of the pure-deflagration scenario can explain the observed structure and the X-ray emission of 3C 397, a multidimensional model that covers the evolutional phase from SN to SNR is needed that includes the asymmetric ejecta distribution, the surrounding environment, the fallback effect, and even impact interaction with the donor.
Observationally, the measurement of the detailed spatial distribution of ejecta clumps and their mass ratios provides important information on the explosion mechanism of SNe Ia, which is enabled by Chandra and XRISM satellites.

While the overall agreement between our computed models and the observed clump abundances  in both the morphology and nucleosynthesis is best for the high central density WD pure deflagration model, there are still some significant error bars (up to 3 $\sigma$) on individual abundance ratios. There is therefore some room in improvement in the hydrodynamical models in terms of inferring the precise parameters of the explosion, both in terms of the WD progenitor properties (in particular its central density), as well as the uncertain ignition parameters. The need for further refinement of the models is particularly evident in the high stellar progenitor WD metallicity (3 $Z_{\odot}$). A similar trend towards higher metallicity was seen in model matches in \citet {yamaguchietal15}, which ultimately derived from the need for higher central density WD model progenitors. We conjecture that a similar effect is at play here, and that further optimization of the WD progenitor structure will narrow down the central density of the progenitor to a slightly higher value than was considered here for our high central density case.

The underlying explosion mechanism is one of the most central questions concerning the origin of the class of SNe Iax events. Some studies have suggested that the structure of SNe Iax are smoothly stratified, which would point towards a detonation mechanism, at least in some brighter events like 2012Z with greater estimated $^{56}$Ni yields \citep {stritzingeretal15, barnaetal17, barnaetal18}. Other investigators have found that a comparison of artificially-mixed profiles of SNe Iax, ranging from smoothly stratified to well-mixed, favors well-mixed ejecta, consistent with the predictions of the failed detonation scenario \citep{Magee_2021}. However, as our model results show, the clumpy distribution observed by \citet {oshiroetal21} disfavors a smoothly stratified ejecta structure that would have originated  for a DDT-powered SN Ia transient for 3C 397.  A pure deflagration origin for 3C 397 opens up a new window onto this central question, with spatially-mapped X-ray observations enabling a direct glimpse into the ejecta.

\section{Conclusions}

In this study, we conducted a detailed analysis of the nucleosynthetic yields and morphology for a wide range of plausible hydrodynamical explosion scenarios of a near-M$_{\rm Ch}$ WD progenitor for SNR 3C 397. 
Our hydrodynamical models of the DDT release sufficient nuclear energy to unbind the whole WD, though  the model ejecta profiles are smoothly stratified, with the most neutronized material located in the central region. Further, the DDT model nucleosynthesis is significantly in disagreement with the observations of 3C 397, with over 9 $\sigma$ deviation in the the ratios of Cr/Fe. While pure deflagration models burn less, and eject lesser still, their ejected mass retains the imprint of the hydrodynamic instabilities from the deflagration phase. Overlaid post-processed particles clearly show  the clustering of neutronized clumps in the pure deflagration ejecta. In particular, our clump analysis of the high central density model reveals clumps rich in Cr/Fe ratios and other neutronized species similar to the observed clump in the southern region of SNR 3C 397. Based on our analysis, we find that SNR 3C 397 most likely originated from the pure deflagration of a high central density of near-M$_{\rm Ch}$ WD progenitor, with an  optical SN Iax transient which gave rise to 3C 397.

SNe Iax are broadly diverse as a class of transients. In marked contrast to the relatively homogeneous class of normal SNe Ia, optical observations demonstrate a remarkable diversity in the spectra and light curves of the class of SNe Iax transients \citep {foleyetal13}. The diversity stems from a number of factors, ranging from the viewing angle dependence of the asymmetric ejecta, to the sensitivity to the inherent stochastic ignition in the core of the near-$M_{\rm Ch}$ WD, to interaction with the non-degenerate donor \citep {bullaetal20}. There may even be entirely distinct explosion channels beyond the near-$M_{\rm Ch}$ scenario which contribute to the SN Iax population, including electron capture SNe \citep {pumoetal09} and the mergers of ONe WDs with CO WDs \citep {kashyapetal18}. Our conclusions  regarding a SN Iax origin of 3C 397 will, of course, need to be confirmed and supported by additional observational and theoretical investigations. ``Reverse engineering'' 3C 397 using simulation models may help us understand in detail the properties of the optical transient which gave rise to the remnant in the context of the near-$M_{\rm Ch}$ channel. 


\section*{Acknowledgements}
R.T.F. acknowledges support from NASA ATP awards 80NSSC18K1013 and 80NSSC22K0630. This work used the Extreme Science and Engineering Discovery Environment (XSEDE) Stampede 2 supercomputer at the University of Texas at Austin’s Texas Advanced Computing Center through allocation TG-AST100038. XSEDE is supported by National Science Foundation grant number ACI-1548562 \citep {townsetal14}. R.T.F. gratefully acknowledges support from the Institute for Advanced Studies and the Heidelberg Institute for Theoretical Studies.

This research has made use of  {FLASH 4.3 \citep {fryxelletal00, dubeyetal12}, the FLASH SN Ia module \citep {townsleyetal16}, Frank Timmes's Adiabatic WD generator code (\url {cococubed.com/code_pages/adiabatic_white_dwarf.shtml}), yt \citep {turketal11}, the Python programming language \citep {vanrossumdeboer91}, Numpy \citep {vanderwaltetal11}, Jupyter/IPython \citep {perezgranger07}, and Matplotlib \citep {hunter07}}.

Data from the astronomical facilities Suzaku \citep {mitsudaetal07} and XMM-Newton \citep {jansenetal01} have been used. 

\section*{Data Availability}

The data underlying this article will be shared on reasonable request to the corresponding author.

\bibliographystyle{mnras}
\bibliography{reference_sne, merged}{}

\begin{table*}
    \centering
    
    \begin{minipage}[c]{0.49\textwidth}
    \resizebox{\textwidth}{!}
    {%
    \begin{tabular}[c]{c|lll}
    \hline
    Species Name & Z = 0 & Z = 1.5 $Z_{\odot}$ & Z = 3.0 $Z_{\odot}$\\
    \hline
     
$^{4}\text{He}$ & $3.03\times10^{-03}$ & $2.36\times10^{-03}$ & $1.77\times10^{-03}$ \\
$^{16}\text{O}$ & $4.28\times10^{-04}$ & $5.07\times10^{-04}$ & $5.10\times10^{-04}$ \\
$^{24}\text{Mg}$ & $2.64\times10^{-05}$ & $4.45\times10^{-06}$ & $2.14\times10^{-06}$ \\
$^{28}\text{Si}$ & $4.75\times10^{-03}$ & $5.05\times10^{-03}$ & $5.14\times10^{-03}$ \\
$^{29}\text{Si}$ & $7.46\times10^{-07}$ & $1.95\times10^{-06}$ & $2.81\times10^{-06}$ \\
$^{30}\text{Si}$ & $1.76\times10^{-06}$ & $3.01\times10^{-06}$ & $6.82\times10^{-06}$ \\
$^{31}\text{P}$ & $9.83\times10^{-07}$ & $1.89\times10^{-06}$ & $3.30\times10^{-06}$ \\
$^{32}\text{S}$ & $3.02\times10^{-03}$ & $2.83\times10^{-03}$ & $2.58\times10^{-03}$ \\
$^{33}\text{S}$ & $1.34\times10^{-07}$ & $2.33\times10^{-06}$ & $3.74\times10^{-06}$ \\
$^{34}\text{S}$ & $8.71\times10^{-07}$ & $1.69\times10^{-05}$ & $3.88\times10^{-05}$ \\
$^{35}\text{Cl}$ & $5.24\times10^{-07}$ & $1.78\times10^{-06}$ & $2.68\times10^{-06}$ \\
$^{36}\text{Ar}$ & $7.00\times10^{-04}$ & $5.85\times10^{-04}$ & $4.99\times10^{-04}$ \\
$^{38}\text{Ar}$ & $1.58\times10^{-07}$ & $1.57\times10^{-05}$ & $3.44\times10^{-05}$ \\
$^{39}\text{K}$ & $5.95\times10^{-07}$ & $2.68\times10^{-06}$ & $3.44\times10^{-06}$ \\
$^{40}\text{Ca}$ & $7.07\times10^{-04}$ & $5.46\times10^{-04}$ & $4.53\times10^{-04}$ \\
$^{42}\text{Ca}$ & $        -         $ & $6.52\times10^{-07}$ & $1.33\times10^{-06}$ \\
$^{44}\text{Ca}$ & $3.57\times10^{-06}$ & $2.22\times10^{-06}$ & $1.34\times10^{-06}$ \\
$^{48}\text{Ti}$ & $2.44\times10^{-05}$ & $1.85\times10^{-05}$ & $1.45\times10^{-05}$ \\
$^{49}\text{Ti}$ & $5.36\times10^{-07}$ & $1.58\times10^{-06}$ & $1.82\times10^{-06}$ \\
$^{51}\text{V}$ & $1.83\times10^{-05}$ & $2.17\times10^{-05}$ & $2.37\times10^{-05}$ \\

    \end{tabular} }
\end{minipage}
\begin{minipage}[c]{0.49\linewidth} 
\resizebox{\textwidth}{!}
    {%
    \begin{tabular}{ c|lll }
    \hline
    Species Name & Z = 0 & Z = 1.5 $Z_{\odot}$ & Z = 3.0 
    $Z_{\odot}$\\
    \hline

$^{50}\text{Cr}$ & $1.19\times10^{-04}$ & $1.30\times10^{-04}$ & $1.45\times10^{-04}$ \\
$^{52}\text{Cr}$ & $1.96\times10^{-03}$ & $1.95\times10^{-03}$ & $1.98\times10^{-03}$ \\
$^{54}\text{Cr}$ & $9.48\times10^{-07}$ & $1.12\times10^{-06}$ & $1.35\times10^{-06}$ \\
$^{53}\text{Mn}$ & $7.62\times10^{-04}$ & $8.15\times10^{-04}$ & $8.51\times10^{-04}$ \\
$^{55}\text{Mn}$ & $1.45\times10^{-02}$ & $1.54\times10^{-02}$ & $1.62\times10^{-02}$ \\
$^{54}\text{Fe}$ & $1.45\times10^{-01}$ & $1.51\times10^{-01}$ & $1.59\times10^{-01}$ \\
$^{56}\text{Fe}$ & $5.47\times10^{-01}$ & $5.20\times10^{-01}$ & $4.94\times10^{-01}$ \\
$^{57}\text{Fe}$ & $2.88\times10^{-02}$ & $3.12\times10^{-02}$ & $3.27\times10^{-02}$ \\
$^{58}\text{Fe}$ & $2.91\times10^{-05}$ & $3.38\times10^{-05}$ & $3.98\times10^{-05}$ \\
$^{58}\text{Ni}$ & $2.08\times10^{-01}$ & $2.26\times10^{-01}$ & $2.44\times10^{-01}$ \\
$^{59}\text{Ni}$ & $2.82\times10^{-03}$ & $2.94\times10^{-03}$ & $2.99\times10^{-03}$ \\ 
$^{60}\text{Ni}$ & $3.60\times10^{-02}$ & $3.45\times10^{-02}$ & $3.36\times10^{-02}$ \\
$^{61}\text{Ni}$ & $3.39\times10^{-04}$ & $3.45\times10^{-04}$ & $2.97\times10^{-04}$ \\
$^{62}\text{Ni}$ & $2.13\times10^{-03}$ & $2.99\times10^{-03}$ & $3.21\times10^{-03}$ \\
$^{63}\text{Cu}$ & $1.63\times10^{-05}$ & $4.33\times10^{-06}$ & $4.61\times10^{-06}$ \\
$^{65}\text{Cu}$ & $2.38\times10^{-06}$ & $2.24\times10^{-06}$ & $1.59\times10^{-06}$ \\
$^{64}\text{Zn}$ & $1.05\times10^{-04}$ & $3.05\times10^{-05}$ & $1.72\times10^{-05}$ \\
$^{66}\text{Zn}$ & $3.93\times10^{-05}$ & $5.19\times10^{-05}$ & $4.78\times10^{-05}$ \\
$^{70}\text{Ge}$ & $1.12\times10^{-06}$ & $1.23\times10^{-06}$ & $9.08\times10^{-07}$ \\
 & & & \\
    \end{tabular} }
\end{minipage}
\vspace{0.5cm}
        \caption{\label{tab:DDT-HIGHDENS-HIGHOFF_Yields} The table contains information about the stable nucleosynthetic yields for DDT-HIGHDENS-HIGHOFF model in mass fraction. The first column shows the name of the species, the second column is the mass fraction for 0 metallicity, the third column is the mass fraction for 1.5$Z_{\odot}$, and the fourth column shows the mass fraction for 3$Z_{\odot}$. The solar metallicity ($Z_{\odot}$) is taken from \citep{asplundetal09}. Please note that the yields below $1 \times 10^{-7}$ are not mentioned in this table.} 
\label{tab:caption}

\vspace{1cm}

\end{table*}

\begin{table*}
    \centering
    
    \begin{minipage}[c]{0.49\textwidth}
    \resizebox{\textwidth}{!}
    {%
    \begin{tabular}[c]{c|lll}
    \hline
    Species Name & Z = 0 & Z = 1.5 $Z_{\odot}$ & Z = 3.0 $Z_{\odot}$\\
    \hline

$^{4}\text{He}$ & $6.78\times10^{-06}$ & $3.33\times10^{-07}$ & - \\
$^{12}\text{C}$ & $1.48\times10^{-01}$ & $1.45\times10^{-01}$ & $1.42\times10^{-01}$ \\
$^{14}\text{N}$ & $1.06\times10^{-07}$ & - & - \\
$^{16}\text{O}$ & $3.79\times10^{-01}$ & $3.72\times10^{-01}$ & $3.65\times10^{-01}$ \\
$^{20}\text{Ne}$ & $5.67\times10^{-03}$ & $4.74\times10^{-03}$ & $4.17\times10^{-03}$ \\
$^{21}\text{Ne}$ & $1.09\times10^{-07}$ & $5.24\times10^{-06}$ & $8.43\times10^{-06}$ \\
$^{22}\text{Ne}$ & - & $1.03\times10^{-02}$ & $2.07\times10^{-02}$ \\
$^{23}\text{Na}$ & $1.55\times10^{-05}$ & $1.09\times10^{-04}$ & $1.70\times10^{-04}$ \\
$^{24}\text{Mg}$ & $5.06\times10^{-03}$ & $3.26\times10^{-03}$ & $2.20\times10^{-03}$ \\
$^{25}\text{Mg}$ & $3.97\times10^{-07}$ & $1.47\times10^{-04}$ & $2.92\times10^{-04}$ \\
$^{26}\text{Mg}$ & $7.52\times10^{-06}$ & $2.71\times10^{-04}$ & $6.00\times10^{-04}$ \\
$^{27}\text{Al}$ & $2.49\times10^{-05}$ & $3.84\times10^{-04}$ & $4.61\times10^{-04}$ \\
$^{28}\text{Si}$ & $5.58\times10^{-02}$ & $5.42\times10^{-02}$ & $5.31\times10^{-02}$ \\
$^{29}\text{Si}$ & $1.99\times10^{-05}$ & $4.08\times10^{-04}$ & $6.86\times10^{-04}$ \\
$^{30}\text{Si}$ & $7.69\times10^{-06}$ & $4.28\times10^{-04}$ & $8.89\times10^{-04}$ \\
$^{31}\text{P}$ & $1.96\times10^{-05}$ & $1.52\times10^{-04}$ & $2.67\times10^{-04}$ \\
$^{32}\text{S}$ & $2.67\times10^{-02}$ & $2.64\times10^{-02}$ & $2.48\times10^{-02}$ \\
$^{33}\text{S}$ & $1.39\times10^{-05}$ & $8.18\times10^{-05}$ & $1.09\times10^{-04}$ \\
$^{34}\text{S}$ & $5.53\times10^{-07}$ & $5.60\times10^{-04}$ & $1.20\times10^{-03}$ \\
$^{36}\text{S}$ & - & - & $2.57\times10^{-07}$ \\
$^{35}\text{Cl}$ & $8.37\times10^{-07}$ & $5.33\times10^{-05}$ & $8.04\times10^{-05}$ \\
$^{37}\text{Cl}$ & $2.19\times10^{-06}$ & $1.37\times10^{-05}$ & $1.88\times10^{-05}$ \\
$^{36}\text{Ar}$ & $4.84\times10^{-03}$ & $4.68\times10^{-03}$ & $4.23\times10^{-03}$ \\
$^{38}\text{Ar}$ & $5.23\times10^{-07}$ & $5.54\times10^{-04}$ & $1.18\times10^{-03}$ \\
$^{39}\text{K}$ & $8.04\times10^{-07}$ & $6.22\times10^{-05}$ & $9.46\times10^{-05}$ \\
$^{41}\text{K}$ & $4.48\times10^{-07}$ & $3.78\times10^{-06}$ & $5.12\times10^{-06}$ \\
$^{40}\text{Ca}$ & $4.10\times10^{-03}$ & $3.97\times10^{-03}$ & $3.55\times10^{-03}$ \\
$^{42}\text{Ca}$ - & $2.44\times10^{-05}$ & $5.43\times10^{-05}$ \\
$^{44}\text{Ca}$ & $2.52\times10^{-06}$ & $2.50\times10^{-06}$ & $2.31\times10^{-06}$ \\
$^{48}\text{Ca}$ & $2.47\times10^{-04}$ & $2.47\times10^{-04}$ & $2.47\times10^{-04}$ \\
$^{45}\text{Sc}$ & - & $1.52\times10^{-07}$ & $1.98\times10^{-07}$ \\
$^{46}\text{Ti}$ & - & $1.14\times10^{-05}$ & $2.24\times10^{-05}$ \\
$^{47}\text{Ti}$ & - & $2.54\times10^{-07}$ & $5.86\times10^{-07}$ \\
    \end{tabular} }
\end{minipage}
\begin{minipage}[c]{0.49\linewidth} 
\resizebox{\textwidth}{!}
    {%
    \begin{tabular}{ c|lll }
    \hline
    Species Name & Z = 0 & Z = 1.5 $Z_{\odot}$ & Z = 3.0 
    $Z_{\odot}$\\
    \hline
$^{48}\text{Ti}$ & $6.97\times10^{-05}$ & $5.98\times10^{-05}$ & $5.21\times10^{-05}$ \\
$^{49}\text{Ti}$ & $9.30\times10^{-05}$ & $9.56\times10^{-05}$ & $9.66\times10^{-05}$ \\
$^{50}\text{Ti}$ & $4.05\times10^{-03}$ & $4.05\times10^{-03}$ & $4.06\times10^{-03}$ \\
$^{51}\text{V}$ & $2.19\times10^{-04}$ & $2.30\times10^{-04}$ & $2.37\times10^{-04}$ \\
$^{50}\text{Cr}$ & $5.55\times10^{-05}$ & $1.10\times10^{-04}$ & $1.75\times10^{-04}$ \\
$^{52}\text{Cr}$ & $6.68\times10^{-03}$ & $6.47\times10^{-03}$ & $6.32\times10^{-03}$ \\
$^{53}\text{Cr}$ & $8.34\times10^{-04}$ & $9.19\times10^{-04}$ & $9.65\times10^{-04}$ \\
$^{54}\text{Cr}$ & $1.36\times10^{-02}$ & $1.36\times10^{-02}$ & $1.37\times10^{-02}$ \\
$^{55}\text{Mn}$ & $7.13\times10^{-03}$ & $7.89\times10^{-03}$ & $8.41\times10^{-03}$ \\
$^{54}\text{Fe}$ & $5.13\times10^{-02}$ & $5.72\times10^{-02}$ & $6.37\times10^{-02}$ \\
$^{56}\text{Fe}$ & $1.69\times10^{-01}$ & $1.62\times10^{-01}$ & $1.55\times10^{-01}$ \\
$^{57}\text{Fe}$ & $6.82\times10^{-03}$ & $7.17\times10^{-03}$ & $7.35\times10^{-03}$ \\
$^{58}\text{Fe}$ & $2.80\times10^{-02}$ & $2.80\times10^{-02}$ & $2.81\times10^{-02}$ \\
$^{59}\text{Co}$ & $1.25\times10^{-03}$ & $1.25\times10^{-03}$ & $1.26\times10^{-03}$ \\
$^{58}\text{Ni}$ & $4.93\times10^{-02}$ & $5.10\times10^{-02}$ & $5.26\times10^{-02}$ \\
$^{60}\text{Ni}$ & $1.43\times10^{-02}$ & $1.43\times10^{-02}$ & $1.44\times10^{-02}$ \\
$^{61}\text{Ni}$ & $2.32\times10^{-04}$ & $2.30\times10^{-04}$ & $2.31\times10^{-04}$ \\
$^{62}\text{Ni}$ & $9.28\times10^{-03}$ & $9.26\times10^{-03}$ & $9.28\times10^{-03}$ \\
$^{64}\text{Ni}$ & $5.89\times10^{-03}$ & $5.90\times10^{-03}$ & $5.90\times10^{-03}$ \\
$^{63}\text{Cu}$ & $6.59\times10^{-05}$ & $6.60\times10^{-05}$ & $6.60\times10^{-05}$ \\
$^{65}\text{Cu}$ & $2.39\times10^{-05}$ & $2.39\times10^{-05}$ & $2.39\times10^{-05}$ \\
$^{64}\text{Zn}$ & $6.51\times10^{-07}$ & $5.85\times10^{-07}$ & $5.91\times10^{-07}$ \\
$^{66}\text{Zn}$ & $2.41\times10^{-03}$ & $2.41\times10^{-03}$ & $2.41\times10^{-03}$ \\
$^{67}\text{Zn}$ & $2.10\times10^{-05}$ & $2.10\times10^{-05}$ & $2.10\times10^{-05}$ \\
$^{68}\text{Zn}$ & $1.61\times10^{-06}$ & $1.62\times10^{-06}$ & $1.62\times10^{-06}$ \\
$^{70}\text{Zn}$ & $4.96\times10^{-06}$ & $4.97\times10^{-06}$ & $4.97\times10^{-06}$ \\
$^{69}\text{Ga}$ & $1.43\times10^{-06}$ & $1.44\times10^{-06}$ & $1.44\times10^{-06}$ \\
$^{72}\text{Ge}$ & $1.05\times10^{-06}$ & $1.05\times10^{-06}$ & $1.05\times10^{-06}$ \\
$^{76}\text{Ge}$ & $1.55\times10^{-06}$ & $1.55\times10^{-06}$ & $1.55\times10^{-06}$ \\
$^{78}\text{Se}$ & $5.10\times10^{-07}$ & $5.11\times10^{-07}$ & $5.11\times10^{-07}$ \\
$^{82}\text{Se}$ & $1.71\times10^{-06}$ & $1.71\times10^{-06}$ & $1.71\times10^{-06}$ \\
$^{86}\text{Kr}$ & $1.87\times10^{-06}$ & $1.87\times10^{-06}$ & $1.87\times10^{-06}$ \\
 & & & \\

    \end{tabular} }
\end{minipage}
\vspace{0.5cm}

\caption{\label{tab:DEF-HIGHDENS-NR-CENTER_Yields}The table contains information about the stable nucleosynthetic yields for DEF-HIGHDENS-NR-CENTER model in mass fraction. The first column shows the name of the species, the second column is the mass fraction for 0 metallicities, the third column is the mass fraction for 1.5$Z_{\odot}$, and the fourth column shows the mass fraction for 3$Z_{\odot}$. The solar metallicity ($Z_{\odot}$) is taken from \citep{asplundetal09}. Please note that the yields below $1 \times 10^{-7}$ are not mentioned in this table.}

\end{table*}

\begin{table*}
    \centering
    
    \begin{minipage}[c]{0.49\textwidth}
    \resizebox{\textwidth}{!}
    {%
    \begin{tabular}[c]{c|lll}
    \hline
    Species Name & Z = 0 & Z = 1.5 $Z_{\odot}$ & Z = 3.0 $Z_{\odot}$\\
    \hline

$^{4}\text{He}$ & $8.52\times10^{-05}$ & $2.45\times10^{-05}$ & $2.11\times10^{-06}$ \\
$^{12}\text{C}$ & $2.55\times10^{-01}$ & $2.50\times10^{-01}$ & $2.45\times10^{-01}$ \\
$^{14}\text{N}$ & $2.65\times10^{-07}$ & - & - \\
$^{15}\text{N}$ & $1.12\times10^{-07}$ & - & - \\
$^{16}\text{O}$ & $3.21\times10^{-01}$ & $3.17\times10^{-01}$ & $3.13\times10^{-01}$ \\
$^{20}\text{Ne}$ & $2.05\times10^{-02}$ & $1.81\times10^{-02}$ & $1.67\times10^{-02}$ \\
$^{21}\text{Ne}$ & $1.78\times10^{-07}$ & $1.40\times10^{-05}$ & $2.28\times10^{-05}$ \\
$^{22}\text{Ne}$ & - & $1.02\times10^{-02}$ & $2.06\times10^{-02}$ \\
$^{23}\text{Na}$ & $6.53\times10^{-05}$ & $3.41\times10^{-04}$ & $5.09\times10^{-04}$ \\
$^{24}\text{Mg}$ & $2.80\times10^{-02}$ & $2.09\times10^{-02}$ & $1.58\times10^{-02}$ \\
$^{25}\text{Mg}$ & $1.63\times10^{-06}$ & $3.52\times10^{-04}$ & $7.31\times10^{-04}$ \\
$^{26}\text{Mg}$ & $3.65\times10^{-05}$ & $7.03\times10^{-04}$ & $1.43\times10^{-03}$ \\
$^{27}\text{Al}$ & $1.44\times10^{-04}$ & $1.61\times10^{-03}$ & $2.08\times10^{-03}$ \\
$^{28}\text{Si}$ & $1.25\times10^{-01}$ & $1.24\times10^{-01}$ & $1.19\times10^{-01}$ \\
$^{29}\text{Si}$ & $6.93\times10^{-05}$ & $1.23\times10^{-03}$ & $2.24\times10^{-03}$ \\
$^{30}\text{Si}$ & $2.52\times10^{-05}$ & $1.59\times10^{-03}$ & $3.28\times10^{-03}$ \\
$^{31}\text{P}$ & $5.83\times10^{-05}$ & $4.52\times10^{-04}$ & $7.91\times10^{-04}$ \\
$^{32}\text{S}$ & $3.83\times10^{-02}$ & $3.75\times10^{-02}$ & $3.67\times10^{-02}$ \\
$^{33}\text{S}$ & $2.16\times10^{-05}$ & $1.70\times10^{-04}$ & $2.41\times10^{-04}$ \\
$^{34}\text{S}$ & $1.55\times10^{-06}$ & $1.44\times10^{-03}$ & $3.03\times10^{-03}$ \\
$^{36}\text{S}$ & $8.89\times10^{-13}$ & - & $3.96\times10^{-07}$ \\
$^{35}\text{Cl}$ & $1.60\times10^{-06}$ & $7.60\times10^{-05}$ & $1.28\times10^{-04}$ \\
$^{37}\text{Cl}$ & $1.33\times10^{-06}$ & $1.32\times10^{-05}$ & $1.99\times10^{-05}$ \\
$^{36}\text{Ar}$ & $5.49\times10^{-03}$ & $5.04\times10^{-03}$ & $4.80\times10^{-03}$ \\
$^{38}\text{Ar}$ & $1.19\times10^{-06}$ & $7.58\times10^{-04}$ & $1.67\times10^{-03}$ \\
$^{39}\text{K}$ & $9.85\times10^{-07}$ & $3.96\times10^{-05}$ & $6.62\times10^{-05}$ \\
$^{41}\text{K}$ & $1.51\times10^{-07}$ & $1.96\times10^{-06}$ & $3.09\times10^{-06}$ \\

    \end{tabular} }
\end{minipage}
\begin{minipage}[c]{0.49\linewidth} 
\resizebox{\textwidth}{!}
    {%
    \begin{tabular}{ c|lll }
    \hline
    Species Name & Z = 0 & Z = 1.5 $Z_{\odot}$ & Z = 3.0 
    $Z_{\odot}$\\
    \hline

$^{40}\text{Ca}$ & $4.28\times10^{-03}$ & $3.84\times10^{-03}$ & $3.62\times10^{-03}$ \\
$^{42}\text{Ca}$ & - & $1.58\times10^{-05}$ & $3.89\times10^{-05}$ \\
$^{44}\text{Ca}$ & $2.52\times10^{-06}$ & $2.25\times10^{-06}$ & $2.13\times10^{-06}$ \\
$^{46}\text{Ti}$ & - & $6.45\times10^{-06}$ & $1.42\times10^{-05}$ \\
$^{47}\text{Ti}$ & - & $1.80\times10^{-07}$ & $4.79\times10^{-07}$ \\
$^{48}\text{Ti}$ & $8.69\times10^{-05}$ & $7.31\times10^{-05}$ & $6.46\times10^{-05}$ \\
$^{49}\text{Ti}$ & $2.27\times10^{-06}$ & $5.76\times10^{-06}$ & $7.17\times10^{-06}$ \\
$^{50}\text{Ti}$ & $5.84\times10^{-06}$ & $5.87\times10^{-06}$ & $5.93\times10^{-06}$ \\
$^{51}\text{V}$ & $9.30\times10^{-06}$ & $2.46\times10^{-05}$ & $3.56\times10^{-05}$ \\
$^{50}\text{Cr}$ & $1.60\times10^{-05}$ & $1.05\times10^{-04}$ & $1.94\times10^{-04}$ \\
$^{52}\text{Cr}$ & $2.44\times10^{-03}$ & $2.17\times10^{-03}$ & $2.11\times10^{-03}$ \\
$^{53}\text{Cr}$ & $1.63\times10^{-04}$ & $2.87\times10^{-04}$ & $3.78\times10^{-04}$ \\
$^{54}\text{Cr}$ & $7.72\times10^{-05}$ & $7.76\times10^{-05}$ & $7.81\times10^{-05}$ \\
$^{55}\text{Mn}$ & $2.22\times10^{-03}$ & $3.32\times10^{-03}$ & $4.22\times10^{-03}$ \\
$^{54}\text{Fe}$ & $1.73\times10^{-02}$ & $2.53\times10^{-02}$ & $3.37\times10^{-02}$ \\
$^{56}\text{Fe}$ & $1.51\times10^{-01}$ & $1.41\times10^{-01}$ & $1.32\times10^{-01}$ \\
$^{57}\text{Fe}$ & $4.31\times10^{-03}$ & $4.99\times10^{-03}$ & $5.35\times10^{-03}$ \\
$^{58}\text{Fe}$ & $3.50\times10^{-04}$ & $3.51\times10^{-04}$ & $3.53\times10^{-04}$ \\
$^{59}\text{Co}$ & $1.52\times10^{-04}$ & $1.40\times10^{-04}$ & $1.30\times10^{-04}$ \\
$^{58}\text{Ni}$ & $2.25\times10^{-02}$ & $2.57\times10^{-02}$ & $2.86\times10^{-02}$ \\
$^{60}\text{Ni}$ & $1.54\times10^{-03}$ & $1.36\times10^{-03}$ & $1.32\times10^{-03}$ \\
$^{61}\text{Ni}$ & $1.71\times10^{-05}$ & $8.63\times10^{-06}$ & $3.98\times10^{-06}$ \\
$^{62}\text{Ni}$ & $2.97\times10^{-04}$ & $2.39\times10^{-04}$ & $1.86\times10^{-04}$ \\
$^{64}\text{Ni}$ & $9.47\times10^{-07}$ & $9.53\times10^{-07}$ & $9.62\times10^{-07}$ \\
$^{63}\text{Cu}$ & $2.49\times10^{-07}$ & $2.08\times10^{-07}$ & $1.71\times10^{-07}$ \\
$^{64}\text{Zn}$ & $8.08\times10^{-07}$ & $1.81\times10^{-07}$ & - \\
$^{66}\text{Zn}$ & $1.12\times10^{-06}$ & $3.61\times10^{-07}$ & - \\

    \end{tabular} }
\end{minipage}
\vspace{0.5cm}
    \caption[The table contains information about the stable nucleosynthetic yields for DEF-STDDENS-NR-CENTER model in mass fraction.]{\label{tab:DEF-STDDENS-NR-CENTER_Yields} The table contains information about the stable nucleosynthetic yields for DEF-STDDENS-NR-CENTER model in mass fraction. The first column shows the name of the species, the second column is the mass fraction for 0 metallicities, the third column is the mass fraction for 1.5$Z_{\odot}$, and the fourth column shows the mass fraction for 3$Z_{\odot}$. The solar metallicity ($Z_{\odot}$) is taken from \citep{asplundetal09}. Please note that the yields below $1 \times 10^{-7}$ are not mentioned in this table.}

\vspace{.5cm}

\end{table*}

\bsp	
\label{lastpage}
\end{document}